\documentclass[twocolumn,showpacs,amsmath,amstex,amssymb,mathfonts,prb]{revtex4-1}
\usepackage{graphicx}

\usepackage{tipa}
\usepackage{bm}

\usepackage{amsmath}
\usepackage{amssymb}
\usepackage{amsthm}
\usepackage{amsfonts}
\usepackage{color}
\usepackage{bbm}
\usepackage{epstopdf}

\newcommand{\norm}[1]{\left\lVert#1\right\rVert}

\begin{document}

\title{Anisotropic Pseudopotential Characterization of Quantum Hall Systems under Tilted Magnetic Field}

\author{Bo Yang$^1$, Ching Hua Lee$^{2,3}$, Chi Zhang$^{4,5}$ and Zi-Xiang Hu$^{6}$}
\affiliation{$^1$Complex Systems Group, Institute of High Performance Computing, A*STAR, Singapore, 138632.\\
$^2$Material Science and Engineering, Institute of High Performance Computing, A*STAR, Singapore, 138632.\\
$^3$Department of Physics, National University of Singapore, Singapore, 117542.\\ 
$^4$International Center for Quantum Materials, Peking University, Beijing, P.R. China, 100871.\\
$^5$Collaborative Innovation Center of Quantum Matter, Beijing, P.R. China, 100871.\\
$^6$Department of Physics, Chongqing University, Chongqing, P.R. China, 401331.}

\pacs{73.43.Lp, 71.10.Pm}

\date{\today}
\begin{abstract}
We analytically derived the effective two-body interaction for a finite thickness quantum Hall system with a harmonic perpendicular confinement and an in-plane magnetic field. The anisotropic effective interaction in the lowest Landau level (LLL) and first Landau level (1LL) are expanded in the basis of the generalized pseudopotentials (PPs), and we analyze how the coefficients of some prominent isotropic and anisotropic PPs depend on the thickness of the sample and the strength of the in-plane magnetic field. We also investigate the stability of the topological quantum Hall states, especially the Laughlin state and its emergent guiding center metric, which we can now compute analytically. An interesting reorientation of the anisotropy direction of the Laughlin state in the 1LL is revealed, and we also discuss various possible experimental ramifications for this quantum Hall system with broken rotational symmetry.
\end{abstract}

\maketitle 
\section{Introduction}
The fractional quantum Hall effect (FQHE) remains as one of the most interesting strongly correlated system for electrons moving in an effective two dimensional manifold~\cite{tsui,prange}. Traditional 2-D electron gas (2DEG) systems fabricated from GaAs/GaAlAs are still the most successful candidates for realizing various FQHE states, including the more exotic Moore-Read states with possible non-Abelian quasiparticles~\cite{MR, Wilczek91, RR}. More recently, FQHE states are also realized in atomically thin 2DEG systems with exposed surfaces (e.g. graphene, ZnO), leading to new ways of influencing the interaction between electrons, as well as interesting physics resulting from additional fermion species~\cite{Kimnature09, Andreinature09}. Theoretically many quantum Hall fluids with fascinating and potentially very useful properties are constructed and predicted~\cite{laughlin,kivelson,simon,lee}, though many of them are very difficult to be realized experimentally. 

Formally, many of the FQH states are best understood via parent model Hamiltonians, consisting of pseudopotentials (PPs) that act as projection operators into certain relative angular momentum sectors of a small cluster of electrons\cite{laughlin,MR,RR}. Interesting phase transitions of the quantum Hall fluids, either between incompressible states or from incompressible to compressible states, can also be understood via the tuning of the linear combination of a small number of PPs, for which the numerical calculation is a powerful tool and the essential physics of such transitions can be understood in a more intuitive way\cite{zlatko,yang1,zhu2017}. From the experimental perspective, the realization of the theoretical predictions of the quantum Hall physics depends on tuning the effective interactions to be close to the corresponding model Hamiltonians. For incompressible topological states, deviation of the physical interaction from the PPs tend to reduce and even close the incompressibility gap, making measurements of many exotic phenomena in the 2+1 spacetime dimension difficult\cite{MR,RR,wen92,haldane08,sarma08}.

Tuning the effective interaction between electrons is the main challenge in realizing exotic topological phases experimentally, since those phases require very special interactions. Experimentally all effective interactions are generally derived from the Coulomb interaction projected into a single Landau level (LL), modified by corrections from LL mixing\cite{DNSheng03,nayak, Ed13, MacDonald13}. For GaAs systems with a finite thickness, one can tune the interaction by varying the thickness of sample~\cite{dassarma08, sternrmp82}. Tilting the magnetic field can also influence the effective interaction by altering the single particle dynamics, and such effect gets stronger when the thickness of the sample increases~\cite{edtilt, xia11, Baldwin13}. For atomically thin samples, the in-plane magnetic field is no longer effective in modifying the interaction between electrons. One can, however, mechanically strain the sample to alter the cyclotron orbits, or to modify the bare Coulomb interaction directly by a conduction layer parallel to the exposed sample surface~\cite{straingraphene, kun13}. The quantum Hall effect can also be potentially realized in a fast rotating cold atom system, in which other ingenious ways can be devised to tune the effective interactions~\cite{Cooper08, qiu11}. 

It is thus important to have a quantitative and more fundamental understanding on how the effective interaction is actually tuned by various choices of ``experimental knobs".  For rotationally invariant physical interactions, such deviations can be quantified by expanding the interactions into the Haldane's PPs\cite{haldane1,haldane2}, which form a complete basis. Anisotropic systems, which include most of the experimental systems mentioned above, are much more non-trivial. Recently, a generalized set of PPs are formulated as a complete basis for any arbitrary two-body interactions, even when rotational symmetry is broken\cite{yang2}. This set of PPs is particularly useful in describing and understanding the interplay between geometry and topology in quantum Hall systems. Armed with such tools, we can look at realistic systems with all possible parameters one can tune experimentally, and to quantitatively understand how the pseudopotential components depend on these parameters.

In this paper, we choose a simple experimental system of a finite thickness 2DEG with a tilted magnetic field as an example, to illustrate how we can completely characterize the effective two-body interactions (which is manifestedly anisotropic) in different LLs in the language of the generalized PPs, and to analyze some of the interesting features of this system. We start by first modeling the finite thickness with a harmonic well and solving the two-body interaction projected in a 2D surface and a single LL analytically. We follow this by expanding the two-body interaction in the set of generalized PPs with different intrinsic metric, so as to explore the tuning of the strength of various PPs by changing the thickness of the sample and the magnitude of the in-plane magnetic field, and to determine the emergent metric of the Laughlin states analytically. 

The organization of the paper is as follows: in Sec.~\ref{exactsolutions}, we present a formal method in solving the effective two-body interaction of the above mentioned system, which does not require the explicit choice of a gauge and leads to a complete analytic expression of the effective interaction. In Sec.~\ref{gpp}, we briefly describe the set of generalized PPs as introduced by Yang et.al\cite{yang2}, and some of their formal properties that were not presented before. In Sec.~\ref{expansion}, we perform the pseudopotential expansion of the effective interaction obtained in Sec.~\ref{exactsolutions} in LLL and 1LL, and analyze the emergent metric of the Laughlin state, which reveals an interesting reorientation of the anisotropy direction in the 1LL, when the strength of the in-plane magnetic field (but not the direction of the field) is varied at different sample thickness. In Sec.~\ref{tlimit} we discuss the capability of the generalized PPs as the theoretical tool in analytically determining the stability and emergent metric of the topological states in the thermodynamic limit, without the ambiguity of the finite-size numerical calculations; we also discuss some important issues from the experimental perspective. In Sec.~\ref{summary}, we summarize our results and discuss possible future developments.

\section{Exact Solution of Effective Interactions}\label{exactsolutions}

For rotationally invariant quantum Hall system confined to a two-dimensional sample with finite thickness, it is well known that the Landau level spectrum and the effective two-body interaction within a single LL can be solved exactly if we model the confining potential in the $z$ direction as a harmonic well. When rotational symmetry is broken by a parallel magnetic field, the LL spectrum can still be solved exactly\cite{maan,chakraborty}, and an analytic expression of the effective interaction can be obtained up to an integral\cite{zlatko}. Here, we present such solutions in a formal way, and in particular show that they can be obtained from the single particle Hamiltonian \emph{without} explicitly picking a gauge for the external magnetic field. We also show that the integral can be performed analytically, leading to more efficient numerical calculations in the following sections. Small parameter expansions in the limit of small sample thickness and small in-plane magnetic field are also performed, to show explicitly the deviation of the approximation of an anisotropic effective mass tensor for such quantum Hall systems, as is sometimes assumed in the literature\cite{yang1}. 

The choice of a harmonic confining potential along the perpendicular direction is due to its technical convenience and good approximation for the realistic systems. Without the in-plane magnetic field, the subband energy levels arising from the harmonic confinement are equally spaced. For detailed and quantitative studies of specific experimental systems, more realistic confinement potentials are needed, and several different models are well studied in the literature for different hetero-junctions\cite{peterson08}. Physically, they lead to subband energies that are no longer equally spaced, and the detailed effective interactions have to be computed numerically instead of analytically. On the other hand, as long as the subband energies are large as compared to the projected Coulomb interactions and we carefully keep track of level crossings when the in-plane magnetic field is introduced, all our methodologies described in this paper can be similarly implemented; different confining potentials generally just introduce different effective length scales along the perpendicular directions\cite{zlatko}, and our results can be generally applied as very good approximations at least within the first few Landau levels. For small subband energy gaps, Landau level and subband level mixing will be more important, leading to effective three-body interactions (or involving more bodies) that are essential for non-abelian FQH states; for this part we will study in details elsewhere.

This section is included here for completeness and its pedagogical interest. Readers not interested in the technical aspects of solving the single particle Hamiltonian can skip this section without affecting the understanding of the rest of the paper.

\subsection{Decoupling and Bogoliubov Transformation}
 
The tilted magnetic field is given by a strong magnetic field applied perpendicular to the sample in the $z$-direction, and an additional component of the in-plane magnetic field. Without loss of generality, we take the in-plane magnetic field to be $B_x$. The single particle Hamiltonian is then given by:
\begin{eqnarray}\label{h1}
H&=&\frac{1}{2m}\left(\left(P_x+eA_x\right)^2+\left(P_y+eA_y\right)^2+\left(P_z+eA_z\right)^2\right)\nonumber\\
&&+\frac{1}{2}m\omega_0^2z^2
\end{eqnarray}
where the second line gives the harmonic potential confinement along z-axis. The canonical momentum can be defined as $\pi_i=P_i+eA_i$, with $i=1,2,3$ along directions $x,y,z$. We also define $\pi_4=m\omega_0z$. Thus Eq.(\ref{h1}) can be written as
\begin{eqnarray}\label{h2}
H=\frac{1}{2m}\left(\pi_1^2+\pi_2^2+\pi_3^2+\pi_4^2\right),
\end{eqnarray}
with the following commutation relations:
\begin{eqnarray}\label{commutation}
&&[\pi_1,\pi_2]=-i\ell_{B_z}^{-2},\nonumber\\
&&[\pi_2,\pi_3]=-i\ell_{B_x}^{-2},\nonumber\\
&&[\pi_3,\pi_1]=[\pi_1,\pi_4]=[\pi_2,\pi_4]=0,\nonumber\\
&&[\pi_3,\pi_4]=-i\ell_0^{-2},
\end{eqnarray}
where the three length scales are given by $\ell_{B_z}=1/\sqrt{eB_z},\ell_{B_x}=1/\sqrt{eB_x},\ell_0=1/\sqrt{m\omega_0}$, with the last one giving the characteristic width of the harmonic well. We can thus define two sets of coupled harmonic oscillators as follows
\begin{eqnarray}\label{ho1}
&&a=\frac{1}{\sqrt{2}}\ell_{B_z}\left(\pi_1-i\pi_2\right),\qquad b=\frac{1}{\sqrt{2}}\ell_0\left(\pi_3-i\pi_4\right),\nonumber\\
&&[a,a^\dagger]=[b,b^\dagger]=1,\nonumber\\
&&[a,b]=[a,b^\dagger]=-\frac{1}{2}\frac{\ell_{B_z}\ell_0}{\ell_{B_x}^{2}}.
\end{eqnarray}
The Hamiltonian can now be rewritten as 
\begin{eqnarray}\label{h3}
H=\frac{1}{2ml_{B_z}^2}\left(a^\dagger a+aa^\dagger\right)+\frac{1}{2ml_0^2}\left(b^\dagger b+bb^\dagger\right).
\end{eqnarray}
One should note here that we have not picked any particular gauge for the external magnetic field. The simple-looking Hamiltonian belies the fact that the two sets of ladder operators are \emph{coupled}. The first step is to decouple the two sets of the harmonic oscillators by taking $a=\alpha+\frac{\ell_{B_z}\ell_0}{2\ell_{B_x}^2}\left(b^\dagger-b\right)$, with $[\alpha,\alpha^\dagger]=1,[\alpha,b]=[\alpha,b^\dagger]=0$, which leads to:
\begin{eqnarray}\label{h4}
H&=&\frac{\omega_z}{2}\left(\alpha^\dagger\alpha+\alpha\alpha^\dagger\right)+\frac{1}{2}\left(\omega_0+\frac{\omega_x^2}{2\omega_0}\right)\left(b^\dagger b+bb^\dagger\right)\nonumber\\
&&-\frac{\omega_x^2}{4\omega_0}\left(b^\dagger b^\dagger +bb\right)\nonumber\\
&&+\frac{\omega_x}{2}\sqrt{\frac{\omega_z}{\omega_0}}\left(\alpha^\dagger b^\dagger+\alpha b-\alpha^\dagger b-\alpha b^\dagger\right)\\
\omega_z&=&\frac{1}{m\ell_{B_z}^2},\quad\omega_x=\frac{1}{m\ell_{B_x}^2}
\end{eqnarray}
We now need to perform a Bogoliubov transformation to bring the Hamiltonian into diagonal form. To do that, we first find the dynamical matrix\cite{xiao} $\mathcal D$ in the basis of $\left(b^\dagger,\alpha^\dagger,b,\alpha\right)$, defined by the commutation $[H,\left(b^\dagger,\alpha^\dagger,b,\alpha\right)^\intercal]=\mathcal D\left(b^\dagger,\alpha^\dagger,b,\alpha\right)^\intercal$, which is given as follows:
\begin{eqnarray}\label{dm}
\mathcal D=\left(\begin{array}{cccc}
\omega_0+\frac{\omega_x^2}{2\omega_0}& -\frac{\omega_x}{2}\sqrt{\frac{\omega_z}{\omega_0}} & -\frac{\omega_x^2}{2\omega_0} & \frac{\omega_x}{2}\sqrt{\frac{\omega_z}{\omega_0}}\\
-\frac{\omega_x}{2}\sqrt{\frac{\omega_z}{\omega_0}} & \omega_z & \frac{\omega_x}{2}\sqrt{\frac{\omega_z}{\omega_0}} & 0\\
\frac{\omega_x^2}{2\omega_0} & -\frac{\omega_x}{2}\sqrt{\frac{\omega_z}{\omega_0}} &-\omega_0-\frac{\omega_x^2}{2\omega_0} &\frac{\omega_x}{2}\sqrt{\frac{\omega_z}{\omega_0}}\\
-\frac{\omega_x}{2}\sqrt{\frac{\omega_z}{\omega_0}} & 0 & \frac{\omega_x}{2}\sqrt{\frac{\omega_z}{\omega_0}} & -\omega_z\end{array}
\right)
\end{eqnarray}
While $\mathcal D$ is non-symmetric, it still can be diagonalized as $\mathcal D=U\Lambda U^{-1}$ with the diagonal matrix $\Lambda$ and $\Lambda_{11}=\omega_1,\Lambda_{22}=\omega_2,\Lambda_{33}=-\omega_1,\Lambda_{44}=-\omega_2$. Defining $\epsilon_1=\omega_z^2+\omega_0^2+\omega_x^2, \epsilon_2=2\omega_0\omega_z$, we have:
\begin{eqnarray}\label{AB}
\omega_1^2&=&\frac{1}{2}\left(\epsilon_1-\sqrt{\epsilon_1^2-\epsilon_2^2}\right),\label{omega1}\\
\omega_2^2&=&\frac{1}{2}\left(\epsilon_1+\sqrt{\epsilon_1^2-\epsilon_2^2}\right).\label{omega2}
\end{eqnarray}

The eigenvectors of $\mathcal D$ form the column vectors of $U$, and a set of decoupled oscillators $X,X^\dagger$ and $Y,Y^\dagger$ given by $\left(X^\dagger,Y^\dagger,X,Y\right)^\intercal=U^{-1}\left(b^\dagger,\alpha^\dagger,b,\alpha\right)^\intercal$. To clean up the notation we set $\omega_z=1$ without loss of any generality. Clearly $\omega_1<1,\omega_2>1$, and $U^{-1}$ is explicitly given as follows:
\begin{eqnarray}
&&U^{-1}=\frac{1}{2\sqrt{\omega_2^2-\omega_1^2}}\left(\begin{array}{cccc}
U_{1}& U_{2}\\
U_{2}^*& U_{1}^*\end{array}
\right)
\end{eqnarray}
\footnotesize
\begin{eqnarray}
&&U_1=\left(\begin{array}{cccc}
i\left(1+\omega_2\right)\sqrt{\frac{1-\omega_1^2}{\omega_2}} & i\left(1+\omega_1\right)\sqrt{\frac{\omega_2^2-1}{\omega_1}}\\
\left(1+\omega_1\right)\sqrt{\frac{\omega_2^2-1}{\omega_1}}& -\left(1+\omega_2\right)\sqrt{\frac{1-\omega_1^2}{\omega_2}}\end{array}
\right)\\
&&U_2=\left(\begin{array}{cccc}
i\left(\omega_2-1\right)\sqrt{\frac{1-\omega_1^2}{\omega_2}}&i\left(1-\omega_1\right)\sqrt{\frac{\omega_2^2-1}{\omega_1}}\\
-\left(1-\omega_1\right)\sqrt{\frac{\omega_2^2-1}{\omega_1}} & \left(\omega_2-1\right)\sqrt{\frac{1-\omega_1^2}{\omega_2}}\end{array}
\right)
\end{eqnarray}
\normalsize
We thus have $[X,X^\dagger]=[Y,Y^\dagger]=1,[X,Y]=[X,Y^\dagger]=0$, and Eq.(\ref{h4}) lead to
\begin{eqnarray}\label{h5}
H=\frac{\omega_1}{2}\left(X^\dagger X+XX^\dagger\right)+\frac{\omega_2}{2}\left(Y^\dagger Y+YY^\dagger\right)
\end{eqnarray}
The single particle Hilbert space is thus built from these two sets of decoupled ladder operators, and the LLs are now indexed by two integers 
$$|m,n\rangle=\frac{1}{\sqrt{m!n!}}\left(X^\dagger\right)^m\left(Y^\dagger\right)^n|0\rangle,$$ where $|0\rangle$ is the vacuum state. In the limit of $\omega_x\rightarrow 0$, $\left(X^\dagger, X\right)$ raises and lowers in-plane LLs, while $\left(Y^\dagger,Y\right)$ raises and lowers the harmonic modes along z-axis (or the subbands), if $\omega_0>\omega_z$. If $\omega_0<\omega_z$, the role of $X$ and $Y$ are reversed. For non-zero $\omega_x$, the in-plane magnetic field mixes the in-plane LLs and the subbands, but $\left(X^\dagger, X\right)$ always give the energy level spacing smaller than that of $\left(Y^\dagger Y\right)$. Thus for non-zero in-plane magnetic field, we treat all energy levels indexed by non-negative integers $m, n$ as ``generalized" LLs.

\subsection{Single-particle Form Factors}
To compute the form factor that enters the effective two-body interaction in a single LL, we now look at the full density-density interaction Hamiltonian with a bare Coulomb interaction
\begin{eqnarray}\label{hint}
H_{\text{int}}=\int d^3qV_{\vec q}\rho_q\rho_{-q},
\end{eqnarray}
where $V_{\vec q}=1/|q|^2$ is the Fourier components of the 3D Coulomb interaction, and $\rho_q=\sum_ie^{i\vec q\cdot r}$ is the density operator. Let us define the usual cyclotron coordinates $\tilde R^{1,2}$ and the guiding center coordinates $\bar R^{1,2}$ as follows:
\begin{eqnarray}\label{coordinates}
&&\tilde R^a=\ell_{B_z}^2\epsilon^{ab}\pi_b, \bar R^a=r^a-\tilde R^a, a=1,2,\nonumber\\
&&[\tilde R^a,\tilde R^b]=-[\bar R^a,\bar R^b]=-i\ell_{B_z}^2\epsilon^{ab},\nonumber\\
&&[\tilde R^a,\bar R^b]=0.
\end{eqnarray}
The part relevant to the Landau level form factor is thus given by
\begin{eqnarray}
&F_{m,n}&\left(\vec q,q_3\right)=\langle m,n|e^{i\left(q_1\tilde R^1+q_2\tilde R^2+q_3r^3\right)}|m,n\rangle\nonumber\\
&=&\langle m,n|e^{PX-P^*X^\dagger+QY-Q^*Y^\dagger}|m,n\rangle\nonumber\\
&=&e^{-\frac{1}{2}\left(PP^*+QQ^*\right)}\mathcal L_m\left(PP^*\right)\mathcal L_n\left(QQ^*\right)
\end{eqnarray}
\begin{eqnarray}
P&=&-i\sqrt{\frac{1-\omega_1^2}{2\omega_1\left(\omega_2^2-\omega_1^2\right)}}q_3\nonumber\\
&&-i\sqrt{\frac{\omega_1\left(\omega_2^2-1\right)}{4\left(\omega_2^2-\omega_1^2\right)}}\left(\left(1+\omega_1\right)q_+-\left(1-\omega_1\right)q_-\right)\\
Q&=&-\sqrt{\frac{\omega_2^2-1}{2\omega_2\left(\omega_2^2-\omega_1^2\right)}}q_3\nonumber\\
&&+\sqrt{\frac{\omega_2\left(1-\omega_1^2\right)}{4\left(\omega_2^2-\omega_1^2\right)}}\left(\left(1+\omega_2\right)q_++\left(\omega_2-1\right)q_-\right)
\end{eqnarray}
where $q_+=\frac{1}{\sqrt{2}}\left(q_1+iq_2\right), q_-=q^*_+, \vec q=(q_1, q_2)$. To get the form factor explicitly it is useful to define the following characteristic functions:
\begin{eqnarray}\label{function}
f_1\left(x,y\right)&=&\frac{\left(x+1\right)\left(y-1\right)\sqrt x}{2\left(y-x\right)},\nonumber\\
f_2\left(x,y\right)&=&\frac{\left(x-1\right)\left(y-1\right)\sqrt x}{4\left(y-x\right)},\nonumber\\
f_3\left(x,y\right)&=&\frac{\sqrt{x\left(1-x\right)\left(y-1\right)}}{\sqrt 2\left(y-x\right)},\nonumber\\
f_4\left(x,y\right)&=&\frac{1-x}{2\sqrt x\left(y-x\right)}.
\end{eqnarray}
Taking $f_i^{12}=f_i\left(\omega_1^2,\omega_2^2\right), f_i^{21}=f_i\left(\omega_2^2,\omega_1^2\right)$ with $i=1,2,3,4$, and $\tilde q^{ij}=f_1^{ij}q_+q_-+f_2^{ij}\left(q_+^2+q_-^{2}\right)+f_3^{ij}q_3\left(q_++q_-\right)+f_4^{ij}q_3^2$, the form factor and the effective two-body interaction are thus given by
\begin{eqnarray}\label{finalf}
&&F_{m,n}\left(\vec q,q_3\right)=e^{-\frac{1}{2}\left(\tilde q^{12}+\tilde q^{21}\right)}\mathcal L_m\left(\tilde q^{12}\right)\mathcal L_n\left(\tilde q^{21}\right)\label{fmn}\\
&&V_{\vec q}^{\left(mn\right)}=\int_{-\infty}^{\infty}dq_3\frac{1}{2q_+q_-+q_3^2}|F_{mn}(\vec q,q_3)|^2\label{effective2d}
\end{eqnarray}

\subsection{Exact Expressions for Form Factors and Small Parameter Expansion}
\subsubsection{Form Factor for the LLL}
With an in-plane parallel magnetic field, rotational symmetry is broken, as can be seen explicitly with the appearance of $q_+^2, q_-^{2}$ and $\left(q_++q_-\right)$ terms in $\tilde q^{ij}$. In the literature, the effect of the in-plane magnetic field is sometimes modeled by a transformation of the effective mass tensor, such that the form factor in Eq.(\ref{finalf}) is approximately characterized by a \emph{single} metric that is stretched and/or rotated. With the exact solution below, we can put this approximation into perspective. One should first note that Eq.(\ref{effective2d}) can be integrated exactly in the LLL (i.e. $m=n=0$), by judiciously solving a set of coupled partial differential equations. The result is as follows:
\begin{eqnarray}\label{v00}
&&V_{\vec q}^{\left(00\right)}=\frac{1}{|q|}e^{-G_1\left(q_+,q_-\right)}G_2\left(q_+,q_-\right)\\
&&G_1\left(q_+,q_-\right)=\left(f_1^{12}+f_1^{21}\right)q_+q_-+\left(f_2^{12}+f_2^{21}\right)\left(q_+^2+q_-^{2}\right)\nonumber\\ &&\\
&&G_2\left(q_+,q_-\right)=\pi\cos F_1e^{F_2}\nonumber\\
&&\qquad -i\sqrt\pi e^{\frac{F_1^2}{4F_2}}\left(\mathcal D\left(\frac{F_1-2iF_2}{2\sqrt{F_2}}\right)-\mathcal D\left(\frac{F_1+2iF_2}{2\sqrt{F_2}}\right)\right)\nonumber\\ &&\\
&&F_1=|q|\left(q_++q_-\right)\left(f_3^{12}+f_3^{21}\right)\\
&&F_2=|q|^2\left(f_4^{12}+f_4^{21}\right)
\end{eqnarray}
Here $G_1\left(q_+,q_-\right)$ clearly gives a squeezed metric in the momentum space for the form factor, while $G_2\left(q_+,q_-\right)$ gives additional corrections, in which $\mathcal D\left(x\right)$ is the Dawson integral\cite{handbook}. In the limit of infinitesimal sample thickness $\omega_0\rightarrow\infty$, $G_1\left(q_+,q_-\right)\rightarrow \frac{1}{2}\left(q_1^2+q_2^2\right)$ and $G_2\left(q_+,q_-\right)\rightarrow 1$. Form factors for higher LLs can also be integrated, though the final expression is complicated.

It is thus clear that, at least in the LLL, approximating a tilted magnetic field system with a deformed metric in the single particle form factor is valid up to a certain degree, with corrections that can be explicitly calculated. For higher LLs, on the other hand, both the exponential part and the Laguerre polynomial part  in Eq.(\ref{fmn}) have very different momentum dependence, especially for relatively small $\omega_0$ (i.e. relatively large sample thickness), and using one deformed metric to characterize the form factor will not be a good approximation in general. This shall be explicitly illustrated below.

\subsubsection{Thin sample limit}

In thin limit of vanishing thickness $l_0$, $\omega_0=\frac1{ml_0^2}\rightarrow \infty$, and any anisotropic effect induced by in-plane magnetic field is small. To second order in $\omega_z/\omega_0$ and $\omega_x/\omega_0$, we have
\begin{align}
\omega_1&\sim \omega_z\left(1-\frac{\omega_x^2}{2\omega_0^2}\right)\\
\omega_2&\sim \omega_0\left(1+\frac{\omega_x^2}{2\omega_0^2}\right).
\end{align}
In other words, a small but finite thickness decreases the energy for the cyclotron excitations by allowing slightly non-perpendicular orbits, but also increases the energy for the confining potential eigenmodes. 
These modified energies in turn modify the effective ``momenta" defining the form factor $F_{m,n}(\vec q,q_z)$ (Eq. \ref{effective2d}):
\begin{align}
\tilde q^{21}&\sim \frac{(q_z\omega_z-q_x\omega_x)^2}{2\omega_0\omega_z}\\
\tilde q^{12}&\sim \frac1{2}(q_x^2+q_y^2)+\frac{\omega_x^2}{4\omega_0^2}\left(\frac{4\omega_z}{\omega_x}q_xq_z-(3q_x^2+q_y^2)\right)
\label{q12q21}
\end{align}
The quantity $\tilde q^{21}$, which affects the form factor dependence on the confining potential energy levels via $\mathcal L_n\left(\tilde q^{21}\right)$, vanishes in the zero thickness limit. However, $\tilde q^{12}$, which affects the form factor dependence on the cyclotron levels, reduces to the untilted expression $\frac1{2}(q_x^2+q_y^2)$ in the zero thickness limit, even when the tilt is nonzero.

To lowest nontrivial order in $1/\omega_0$ (and setting $\omega_z=1$ again), the effective two-body interactions are given by
\begin{eqnarray}
& &V^{(mn)}_{\vec q} = e^{-\frac{|\vec q|^2}{2}} \nonumber \\
           & &\int_{-\infty}^\infty \frac{e^{-(q_z-q_x\omega_x)^2/(2\omega_0)}}{|\vec q|^2+q_z^2}\mathcal L_m\left(\tilde q^{12}\right)\mathcal L_n\left(\tilde q^{21}\right)dq_z
\label{Vmnomega_0}
\end{eqnarray}
where $\vec q=(q_x,q_y)$, and $\tilde q^{12}$ and $\tilde q^{21}$ are given by Eq. (\ref{q12q21}). $V^{(mn)}_{\vec q}$ can be evaluated in terms of the auxiliary function $\Gamma(Q,\gamma)$ and its $\gamma$-derivatives, where
\begin{eqnarray}
&&\Gamma(Q,\gamma)\notag\\
&=& \int_{-\infty}^\infty \frac{e^{-(q_z-\gamma)^2}}{Q^2+q_z^2}dq_z\notag\\
&=&\frac{\pi e^{(\gamma+iQ)^2}\left(1+e^{4i\gamma Q}(1-\text{Erf}(i\gamma+Q))+ \text{Erf}(i\gamma-Q)\right)}{2Q}\notag\\
\label{Gamma}
\end{eqnarray}
with $\text{Erf}(z)=\frac{2}{\sqrt{\pi}}\int_0^ze^{-t^2}dt$ being the Error function. 

Keeping the lowest two orders of $1/\omega_0$, we find the effective LLL potential
\begin{eqnarray}
V^{(00)}_{\vec q}&\sim& \pi \frac{e^{-|\vec q|^2/2}}{|\vec q|}\left(1-\sqrt{\frac{2}{\pi\omega_0}}|\vec q|+\frac{|\vec q|^2-2q_x^2\tan^2\phi}{2\omega_0}\right)\notag\\
&\sim& \pi\frac{e^{-\frac1{2}g^{(00)}_{\mu\nu}q^\mu q^\nu}}{|\vec q|}-e^{-|\vec q|^2/2}\sqrt{\frac{2\pi}{\omega_0}}
\label{V00thin}
\end{eqnarray}
where $\phi =\tan^{-1}\frac{\omega_x}{\omega_z}$ is the tilt angle. In this form, $V^{(00)}_{\vec q}$ is explicitly given by a Coulomb interaction regulated by a Gaussian with effective metric $g^{(00)}$. Up to a subleading $\mathcal{O}(\omega_0^{-1/2})$ residual term 
, $V^{(00)}_{\vec q}$ is indeed well approximated by a single effective metric given by $g^{(00)}=\text{diag}\left(1-\frac{1-2\tan^2\phi}{\omega_0},1-\frac1{\omega_0}\right)$. This metric is proportional to one with deformation parameter $\frac{\tan^2\phi}{\omega_0}$.  

Similarly, we also obtain for the 1LL
\begin{eqnarray}
&&V^{(10)}_{\vec q} \sim \pi \frac{e^{-|\vec q|^2/2}}{|\vec q|} \notag \\
& &\left(1-\frac{|\vec q|^2}{2}\right)\left(1-\sqrt{\frac{2}{\pi\omega_0}}|\vec q|+\frac{|\vec q|^2-2q_x^2\tan^2\phi}{2\omega_0}\right) \notag\\
&\sim& \left(1-\frac{|\vec q|^2}{2}\right)V^{00}_{\vec q},
\end{eqnarray}
with the same effective metric as that of $V^{(00)}$, but multiplied by the well-known factor of $1-\frac{|\vec q|^2}{2}$. 

Thus for thin samples, a tilted magnetic field induces a squeezing of the effective mass tensor in the LLL, as is the common approximation employed in the literature. However, in this case the effective mass tensor in the first term of Eq.(\ref{V00thin}) is also dilated, and there is an additional subleading Gaussian correction. For the 1LL, a tilted magnetic field results in a complicated effective two-body interaction containing three different metrics even in the thin sample limit, and the squeezed effective mass tensor approximation of the form factor is not very appropriate.

\begin{figure*}[htb]
\includegraphics[width=0.8\linewidth]{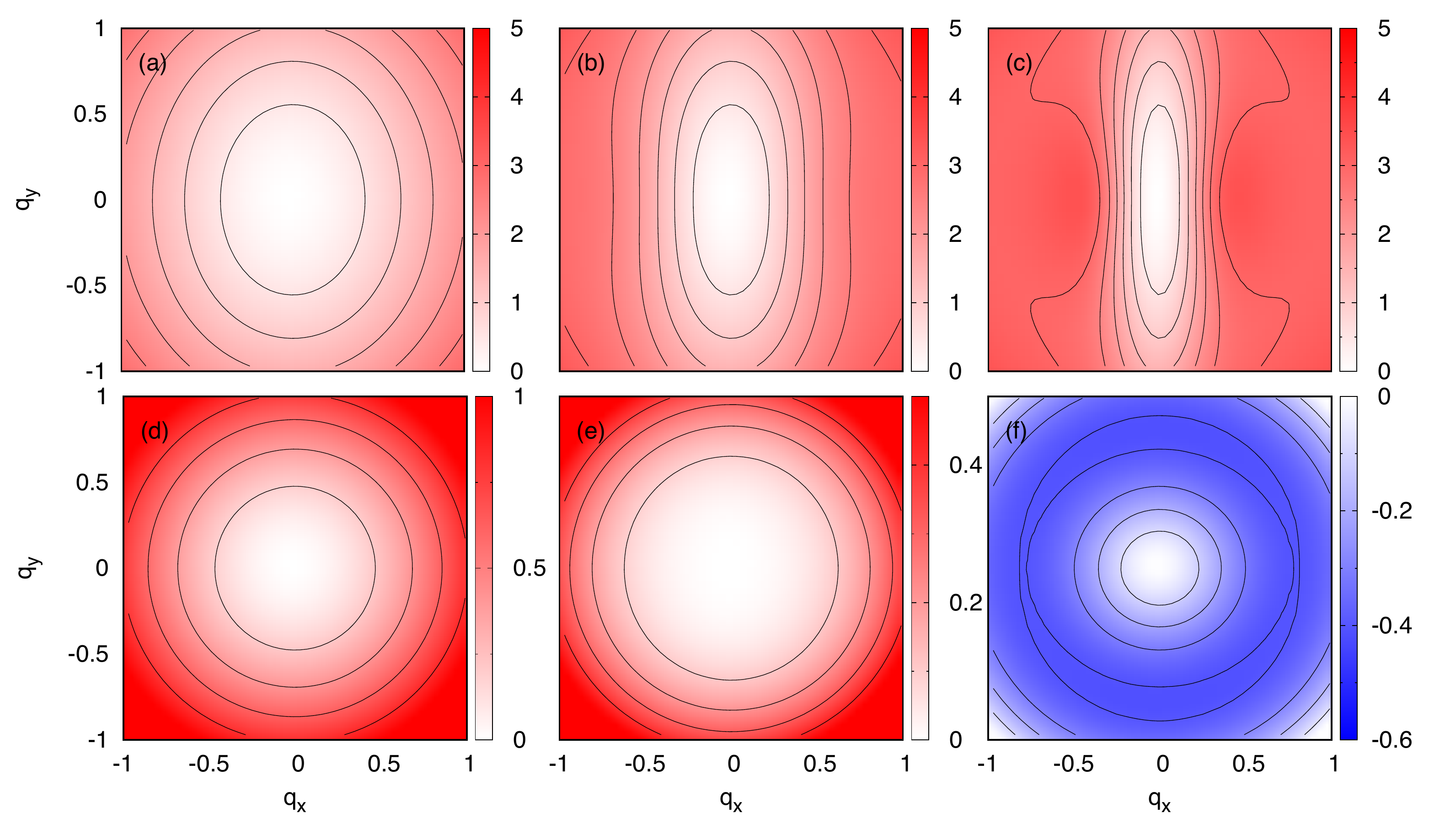}
\caption{Effective metric plots $g^{(00)(\vec q)}_{\mu\nu}q^\mu q^\nu$. Top: Thin limit with $\omega_0=5\omega_z$ and $\omega_x/\omega_z=\tan\phi=2,5,10$. Bottom: Small tilt limit with $\tan\phi=0.2$ and $\omega_0/\omega_z=1,0.5,0.2$. Only in the left column does the effective metric remain approximately constant for $|\vec q|<1$. While tilting the magnetic field primarily induces anisotropy, increasing the thickness primarily modulates the radial dependence of the metric.  }
\label{LLmetric}
\end{figure*}

\subsubsection{Small tilt limit}
We now consider the other limit, the limit of small tilt $\omega_x=\frac{eB_x}{m}\rightarrow 0$. To second order in $\omega_x/\omega_0$ and $\omega_x/\omega_z=\tan\phi$, we have 
\begin{align}
\omega_1&\sim \omega_0\left(1-\frac{\omega_x^2}{2(\omega_z^2-\omega_0^2)}\right)\\
\omega_2&\sim \omega_z\left(1+\frac{\omega_x^2}{2(\omega_z^2-\omega_0^2)}\right),
\end{align}
assuming $\omega_0<\omega_z$. Otherwise, $\omega_1,\omega_2$ shall be switched. These small tilt approximations are valid except when $\omega_z\approx \omega_0$, whereby 
\begin{eqnarray}
\frac{\omega_{1,2}|_{\omega_0= \omega_z}}{\omega_0}&=&\sqrt{1+\left(\frac{\omega_x}{2\omega_z}\right)^2}\mp \frac{\omega_x}{2\omega_z}\notag\\
&\approx & \sec\frac{\phi}{2}\mp \tan \frac{\phi}{2}\approx 1\mp \frac{\phi}{2}
\end{eqnarray}
where $\tan\phi=\frac{\omega_x}{\omega_z}$. For small tilt $\phi$ in the thin regime of $\omega_0>\omega_z$, these lead to (setting $\omega_z=1$ unless otherwise stated)
\begin{align}
\tilde q^{21}&\sim \frac{q_z^2}{2\omega_0}-\frac{q_xq_z\omega_0}{\omega_0^2-1}\tan\phi \\
\tilde q^{12}&\sim \frac1{2}(q_x^2+q_y^2)+\frac{q_xq_z}{\omega^2_0-1}\tan\phi
\label{q12q21b}
\end{align}
To capture the thick sample regime $\omega_0<\omega_z$, one simply exchanges $\tilde q^{21}$ and $\tilde q^{12}$. In the critical regime of $\omega_z\approx \omega_0$, we also have
\begin{align}
\tilde q^{12},\tilde q^{21}|_{\omega_0= \omega_z}&\sim \frac1{4}(q_y^2+(q_x\pm q_z)^2)-\frac1{4}q_x(q_x\pm q_z)\tan\phi
\label{q12q21c}
\end{align}
To lowest nontrivial order in $1/\omega_x$, the effective two-body interactions are now given by
\begin{eqnarray}
V^{(mn)}_{\vec q}&=&e^{\frac{\omega_0q_x^2\omega_x^2}{2(1+\omega_0)^2}}e^{-\frac{|\vec q|^2}{2}} \int_{-\infty}^\infty \frac{e^{-(q_z-q_x\omega_x\omega_0/(1+\omega_0))^2/(2\omega_0)}}{|\vec q|^2+q_z^2}\notag\\
&&\times\mathcal L_m\left(\tilde q^{12}\right)\mathcal L_n\left(\tilde q^{21}\right)dq_z
\label{Vmnomega_b}
\end{eqnarray}
where $\tilde q^{12},\,\tilde q^{21}$ are given by Eq. \ref{q12q21b} or  \ref{q12q21c}. This is the same as Eq. \ref{Vmnomega_0} for the thin limit, if one makes the replacement $\omega_x\rightarrow \omega_x \frac{\omega_0}{1+\omega_0}$, employs the modified expressions for $\tilde q^{12},\,\tilde q^{21}$ and multiplies by an overall factor $e^{\frac{\omega_0q_x^2\omega_x^2}{2(1+\omega_0)^2}}$.

Keeping the lowest few orders of $\omega_x$, we thus find the effective LLL potential
\footnotesize
\begin{eqnarray}
V^{(00)}_{\vec q}&\sim& \pi \frac{e^{-|\vec q|^2/2}}{|\vec q|}\notag\\
&&\times\left(A[z]+\left(\sqrt{\frac{2}{\pi\omega_0}}|\vec q|-A[z]\left(1+\frac{|\vec q|^2}{\omega_0}\right)\right)\frac{q_x^2\tan^2\phi}{\omega_0}\right)\notag\\
\end{eqnarray}
\normalsize
where $z=\frac{|\vec q|}{\sqrt{2\omega_0}}$ and $A[z]=e^{z^2}(1-\text{Erf}[z])\sim 1-\frac{2}{\sqrt{\pi}}z+z^2+O(z^3)$. For small $\vec q$, we obtain for this small tilt limit an identical effective metric as in the thin limit, which is given by Eq. \ref{V00thin} and 
\[ g^{(00)}=\text{diag}\left(1-\frac{1-2\tan^2\phi}{\omega_0},1-\frac1{\omega_0}\right)\]

\subsubsection{Small $\vec q$ limit}
For any tilt and thickness, it is also instructive to see how the potential behaves for small $\vec q$. To fourth order in $\vec q$, $|\vec q|e^{|\vec q|^2/2}V^{(00)}/\pi$ is given by
\footnotesize
\begin{eqnarray}
&&1-\frac{\sqrt{\frac{2}{\pi }} |\vec q|}{\sqrt{\omega_0}}-\frac{|\vec q|^2 \left(-1+\omega_x^2+\omega_x^2 \cos [2 \theta ]\right)}{2 \omega_0}\notag\\
&&+\frac{\sqrt{\frac{2}{\pi }} |\vec q|^3 \left(-1+3 \omega_x^2+3 \omega_x^2 \cos [2 \theta ]\right)}{3 \omega_0^{3/2}}\notag\\
&&+\frac{|\vec q|^4 \left(2-12 \omega_x^2+3 \omega_x^4+4 \omega_x^2 \left(-3+\omega_x^2\right) \cos [2 \theta ]+\omega_x^4 \cos [4 \theta ]\right)}{16 \omega_0^2}\notag\\
\end{eqnarray}
\normalsize
where $\theta=\tan^{-1}\frac{q_y}{q_x}$. We observe a qualitative difference between the effects of tilt and thickness: A tilted magnetic field primarily modifies the anisotropy of the effective metric, while finite thickness primarily induces modulations of the metric in the radial $|\vec q|$ direction. This is illustrated in Fig. \ref{LLmetric} for a few typical to extreme cases.

\section{Generalized Two-Body Pseudopontentials}\label{gpp}

The exact solution obtained in Eq.(\ref{effective2d}) implies that to understand the effective interaction for a finite thickness sample with a parallel magnetic field, using a deformed metric for the single particle form factor is generally not enough. We thus use the generalized PPs as a set of complete basis to characterize the interaction. The motivations and basic formalism of the generalized PPs are introduced in Ref.~\onlinecite{yang2}. It is a useful tool for analyzing anisotropic effective interactions and for studying the interplay between geometry and topology of the quantum Hall systems, as we will illustrate in details in the following sections. For completeness, in this section we briefly reproduce the definition of the generalized PPs and discuss some of the formal properties.

The generalized PPs are parametrized by a unimodular metric. Using the complex vector $\omega_a$ satisfying $\epsilon^{ab}\omega_a^*\omega_b=i$ (Einstein's summation rule is assumed), the metric is given by $g_{ab}=\omega_a^*\omega_b+\omega_a\omega_b^*$, with $\det g=1$ and $\omega^a=g^{ab}\omega_b$. The generalized PPs, indexed by two non-negative integers $r,s$, are given as follows:
\begin{eqnarray}
V_{r,s}^{g+}\left(\vec q\right)&=& \lambda_s \mathcal{N}_{rs} \left(L_r^s\left(|q|^2\right)e^{-\frac{1}{2}|q|^2}q_+^s+c.c\right)\label{g1s}\\
V_{r,s}^{g-}\left(\vec q\right)&=& -i \mathcal{N}_{rs} \left(L_r^s\left(|q|^2\right)e^{-\frac{1}{2}|q|^2}q_+^s-c.c\right)\label{g2s}
\end{eqnarray}
Here, $|q|^2=g^{ab}q_aq_b, q_+=\omega^aq_a$, and $q_-=q_+^*$; $\lambda_s=1/\sqrt{2}$ if $s=0$, and $\lambda_s=1$ if $n\neq 0$. The normalization factors are $\mathcal N_{rs}=\sqrt{2^{s-1}r!/\left(\pi\left(r+s\right)!\right)}$. These PPs are orthogonal:$\int d^2qV_{r,s}^{g\sigma}V_{r',s'}^{g\sigma'}=\delta_{r,r'}\delta_{s,s'}\delta_{\sigma,\sigma'}$, and they form a complete basis for any arbitrary interactions in 2D. The isotropic Haldane's pseudopotentials are special cases of Eq.(\ref{g1s}) with $s=0$, we thus use $V_{r,0}^{g+}$ and $V_r^g$ (as well as $c_{r,0}^{g+}$ and $c_r^g$ in Eq.(\ref{decompose2s})) interchangeably; note also that $V_{r,0}^{g-}=0$. We can thus give the expansion as follows:
\begin{eqnarray}
V_{\vec{q}}&=&\sum_{r,s=0}^\infty \sum_{\sigma=\pm}   c^{g\sigma}_{r,s}V_{r,s}^{g\sigma}(\vec{q}),\label{decompose2s}\\
c^{g\sigma}_{r,s}&=&\int d^2q V_{\vec{q}} V_{r,s}^{g\sigma}\left(\vec q\right).\label{decompose2as}
\end{eqnarray}
where we show explicitly that for the same physical interaction, the coefficients of expansion depend on the metric of choice in the generalized PPs. We can thus tune the metric to vary the coefficients, so as to expose the nature of the interaction in more suitable coordinates. If the norm of $V_{\vec q}$ is integrable over the momentum space, we have the following relationship:
\begin{eqnarray}\label{l2}
\int d^2qV_{\vec q}^2=\sum_{r,s=0}^\infty\sum_{\sigma=\pm}\left(c_{r,s}^{g\sigma}\right)^2
\end{eqnarray}
Thus by treating $\vec c_g$ as a vector with $c_{r,s}^{g\pm}$ as entries, tuning the metric $g$ only changes the direction of this vector without altering its magnitude. In fact, it is evident numerically that we have six vectors defined as follows that conserve their respective magnitude when we vary $g$:
\begin{eqnarray}\label{identity}
&&\norm{c_1^+}^2=\sum_{r,s=0}^\infty \left(c_{2r+1,2s}^{g+}\right)^2,\norm{c_1^-}^2=\sum_{r,s=0}^\infty \left(c_{2r+1,2s}^{g-}\right)^2\nonumber\\
&&\norm{c_2^+}^2=\sum_{r,s=0}^\infty \left(c_{r,2s+1}^{g+}\right)^2,\norm{c_2^-}^2=\sum_{r,s=0}^\infty \left(c_{r,2s+1}^{g-}\right)^2\nonumber\\
&&\norm{c_3^+}^2=\sum_{r,s=0}^\infty \left(c_{2r,2s}^{g+}\right)^2,\norm{c_3^-}^2=\sum_{r,s=0}^\infty \left(c_{2r,2s}^{g-}\right)^2
\end{eqnarray}
For bosons and fermions, only $V_{r,2s}^{g\pm}$ are relevant, and deforming $g$ will not mix $V_{2r,2s}^{g\pm}$ with $V_{2r+1,2s}^{g\pm}$. If $V_{\vec q}$ is rotationally invariant, there exists a metric $g$ (which is not necessarily $\mathbbm{1}_{2\times 2}$) such that all $c_{r,s}^{g\sigma}$ with $s\neq 0$ or $\sigma=-$ vanishes. For anisotropic interactions such as the ones computed in the previous section with samples of finite thickness and non-zero parallel magnetic field, only $V^{g+}_{r,s}$ needs to be considered given that the in-plane magnetic field is along $x$-axis.  

\section{Pseudopotential Description of Effective Interactions}\label{expansion}

From now onwards, the metric superscript for the pseudopotentials are omitted if the lab frame metric $\left(1,0,0,1\right)$ is used. With the analytic expressions of the effective two-body interaction and the definition of the generalized PPs, the evaluation of Eq.(\ref{decompose2as}) with an arbitrarily chosen metric can be done numerically with Mathematica. We will now carry out various types of the expansion of the effective interaction, with the following questions in mind:
\begin{itemize}
\item How will the isotropic PPs be tuned by the sample thickness and parallel magnetic field;
\item What is the scale of the magnitude of the anisotropic PPs as compared to the isotropic PPs;
\item Are there any qualitative difference for the effective interaction between LLL and higher LLs; 
\item How can the stability of the $\nu=1/3$ Laughlin state be accounted for by the metric variation in the PPs;
\end{itemize}
We thus focus on the interactions $V_{\vec q}^{\left(00\right)}$ and $V_{\vec q}^{\left(10\right)}$ as defined in Eq.(\ref{effective2d}), corresponding to the dynamics in LLL and 1LL. 

\subsection{Effective Interaction in the LLL}\label{LLL_description}

\begin{figure}[htb]
\includegraphics[width=\linewidth]{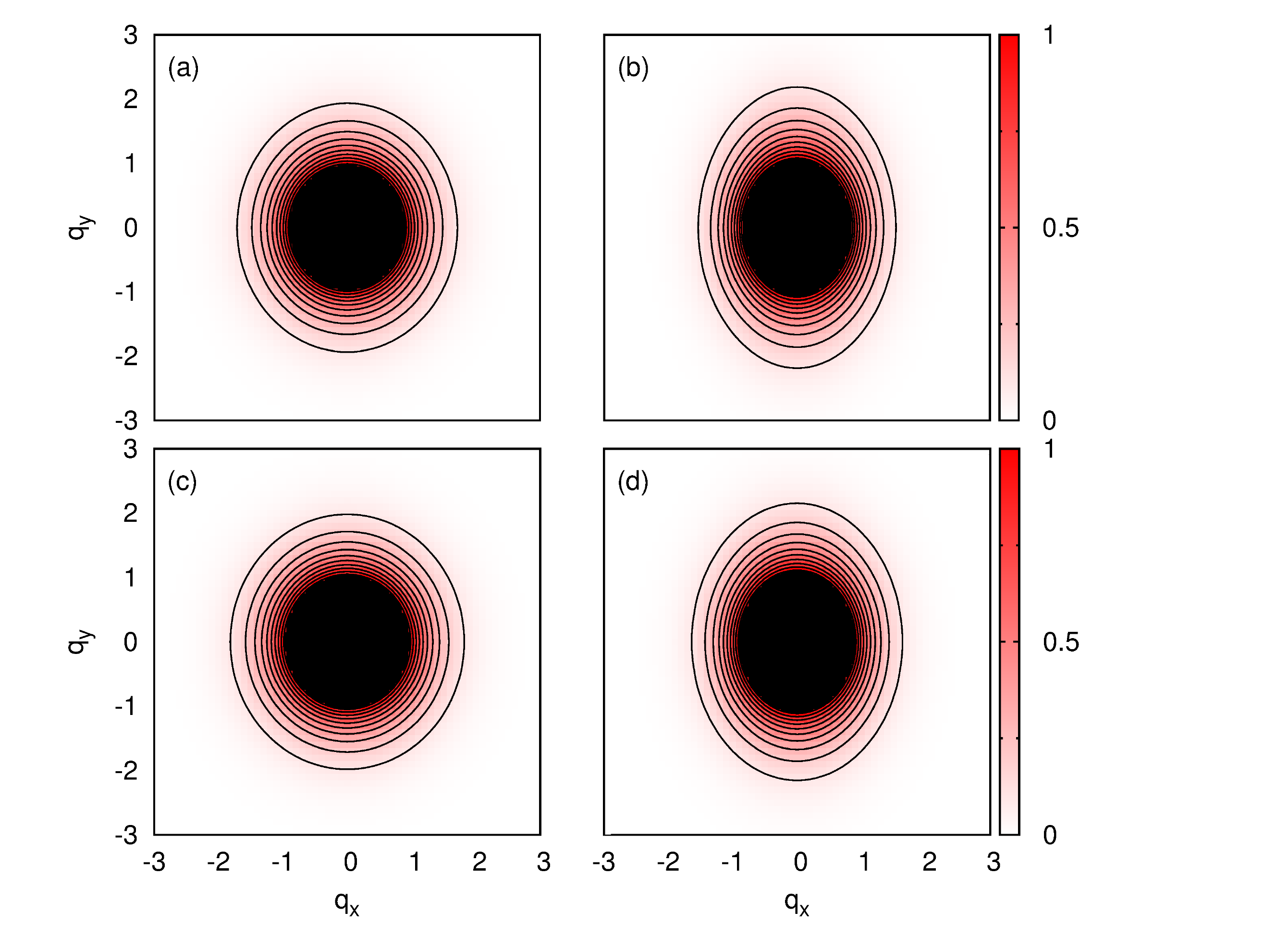}
\caption{The contour plots of $V_{\vec q}^{\left(00\right)}$ at different thickness and parallel magnetic field. a)$\omega_0=0.6,\omega_x=1$; b) $\omega_0=0.6,\omega_x=2$; c) $\omega_0=1.2,\omega_x=1$; d) $\omega_0=1.2,\omega_x=2$. Note that in LLL the contours of the projected interaction are relatively featureless.}
\label{contour_LLL}
\end{figure}
\begin{figure}[htb]
\includegraphics[width=\linewidth]{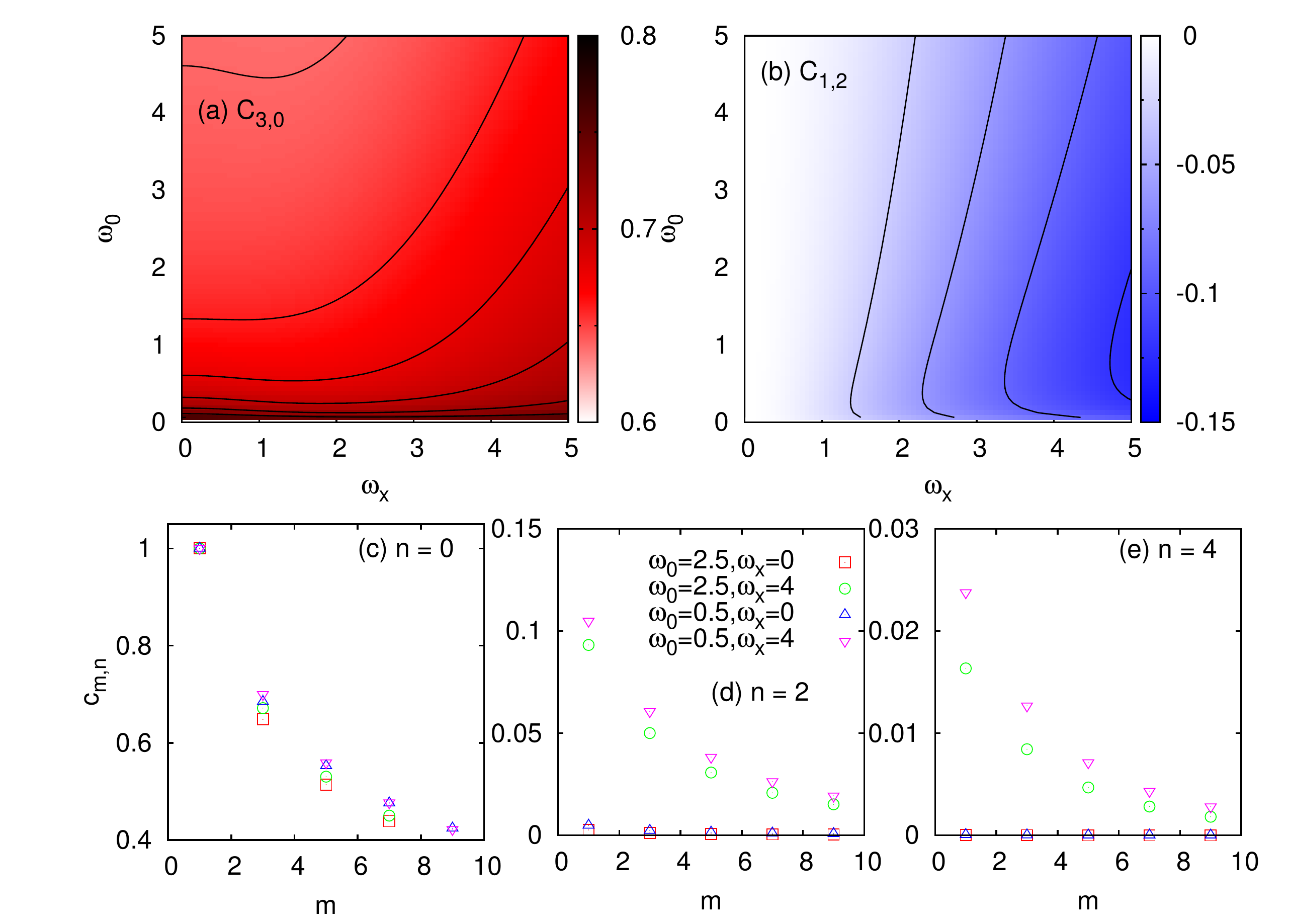}
\caption{a)The contour plot of $c_{3,0}$ (no superscript needed for isotropic PPs); b) The contour plot of $c_{1,2}^+$ (note that $c_{m,n\neq 0}^-$ vanishes); c) The dependence of $c_{m,0}$ on $m$ at different sample thickness and parallel magnetic field; d) The dependence of $c_{m,2}^+$ on $m$ at different sample thickness and parallel magnetic field; e) The dependence of $c_{m,4}^+$ on $m$ at different sample thickness and parallel magnetic field. The coefficient of $V_{1,0}$ (i.e. $c_{1,0}$) is normalized to unity. The metric in the pseudopotentials is (1,0,0,1). c),d),e) share the same legend in d).}
\label{f1}
\end{figure}

 
For a consistent comparison of the relative strength of different pseudopotentials, we normalize the coefficient of $V_1^g$ (note for isotropic PPs we write $V^g_m=V^{g+}_{m0}$) to unity. It is natural to first expand the effective interaction in terms of the generalized PPs with the lab metric, which we set to be $g=\eta=\text{\textit{diag}}(1,1)$. For a wide range of sample thickness and parallel magnetic field, the most important isotropic PP (apart from $V^\eta_1$) is $V^\eta_3$, while the most important anisotropic PP is $V^{\eta+}_{12}$ (the coefficients of $V^{\eta-}_{mn}$ vanishes due to inversion symmetry), and we plot them in Fig. \ref{f1}. 
It is clear that when the thickness of the sample decreases as given by increasing $\omega_0$, the anisotropic part of the effective interaction becomes less important; in the limit of infinitesimal sample thickness, the parallel magnetic field no longer affects the effective interaction. The anisotropic part of the effective interaction is large when the sample is thick and the parallel magnetic field is large, as expected; however, even for really large tilting of the magnetic field, the coefficients of the anisotropic PPs are almost one order of magnitude smaller than the isotropic pseudopotentials. The dominant anisotropic PPs are $V^\eta_{m,2}$, and it is a very good approximation to ignore all $V^\eta_{m,n}$ with $n\ge 4$.

\begin{figure}[htb]
\includegraphics[width=0.8\linewidth]{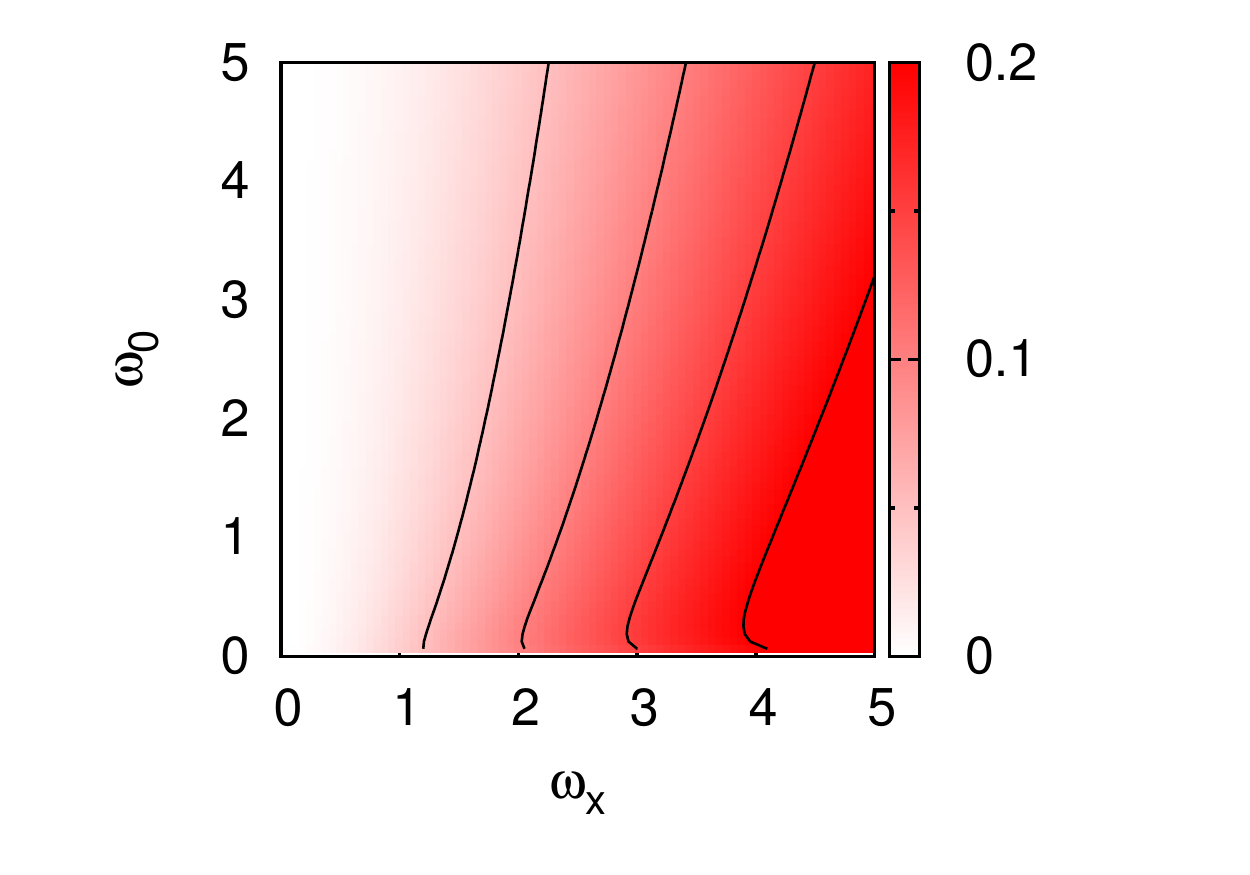}
\caption{The $\theta$ of the metric that maximizes $V^g_1$ in LLL, for such metric $\phi=0$.}
\label{LLLtheta}
\end{figure}

We now look at the metric of the PPs which can serve as a variational parameter. As we illustrated in Eq.(\ref{g1s}) and Eq.(\ref{g2s}), the generalized PPs are parameterized by a unimodular metric g with two degrees of freedom corresponding to stretching and rotation of the metric as follows:
\footnotesize
\begin{eqnarray}\label{g}
g=\left(\begin{array}{ccc}
\cosh 2\theta+\sinh 2\theta\cos 2\phi & \sin 2\phi\sinh 2\theta\\
\sin 2\phi\sinh 2\theta & \cosh 2\theta-\sinh 2\theta\cos 2\phi\end{array}\right)
\end{eqnarray}
\normalsize
where $\theta$ parametrizes the stretching of the metric, and $\phi$ parametrizes the orientation of the stretched metric. The most important PP for the Laughlin state at $\nu=1/3$ is $V^g_1$. In Fig. \ref{LLLtheta} we plot $\theta$ that maximizes the coefficient of $V^g_1$. Since the Laughlin state is the exact ground state of the $V^g_1$, all other PPs in the effective interaction can be treated as perturbations to the Laughlin state. One should note that according to Eq.(\ref{identity}), as long as Eq.(\ref{l2}) is bounded, the metric that maximizes the coefficient of $V^g_1$ also minimizes the sum of the square of the coefficients of all other PPs. Even though the square of $V_{\vec q}^{\left(mn\right)}$ is not integrable, it can be regulated\cite{ippoliti} at small and large $|q|$. We thus expect the Laughlin state with this metric to have the maximal overlap with the true ground state of $V_{\vec q}^{\left(00\right)}$.

It is clear from Fig. \ref{LLLtheta} that $\theta$ generally decreases when the sample thickness decreases. When $\omega_x<1$ or the tilt angle is smaller than $\pi/4$, $\theta$ is generally very small. At fixed tilted angle of the magnetic field, $\theta$ is maximized for some small $\omega_0<1$, due to the interplay of the sample thickness and the magnetic length of the total magnetic field. One should note that small $\omega_0$ does not always mean very thick sample; since it is measured against $\omega_z=1$, it could also imply very strong perpendicular magnetic field. There is also an interesting relationship that dictates the metric that maximizes $V^g_1$ also makes $V^{g+}_{12}$ vanishes, which can be easily checked from the PP definitions:
 \begin{eqnarray}\label{relationship}
 \partial_\theta V_{1}^g=-\frac{\sqrt{3}}{2} \left( \cos\phi V_{1,2}^{g+} - \sin\phi V_{1,2}^{g-}\right).
 \end{eqnarray}
and this can be easily verified with Eq.(\ref{g1s}) and Eq.(\ref{g2s}). In the LLL, by changing from the metric of the lab frame to the metric that maximizes the coefficient of $V_1^g$ in the pseudopotential expansion, the coefficients of the isotropic PPs ($V_{m0}$ with $m\ge 3$) decreases. The coefficients of the anisotropic PPs ($V^g_{mn}$ with $m\ge 3, n\ne 0$) increases, though to a lesser degree, and $V^g_{12}$ vanishes completely. For the same effective interaction, using different metric in the pseudopotential expansion is equivalent. However, if the ground state is gapped and adiabatically connected to the Laughlin state at $\nu=1/3$ filling factor, the metric that maximizes the coefficient of $V^g_1$ is more appropriate in understanding how pseudopotentials other than $V^g_1$ perturbs the ground states and excites the quasiparticles. Such metric can thus be considered as the emergent intrinsic guiding center metric of the many-body Laughlin state.

\subsection{Effective Interactions in the 1LL}\label{1LL_description}

\begin{figure}[htb]
\includegraphics[width=\linewidth]{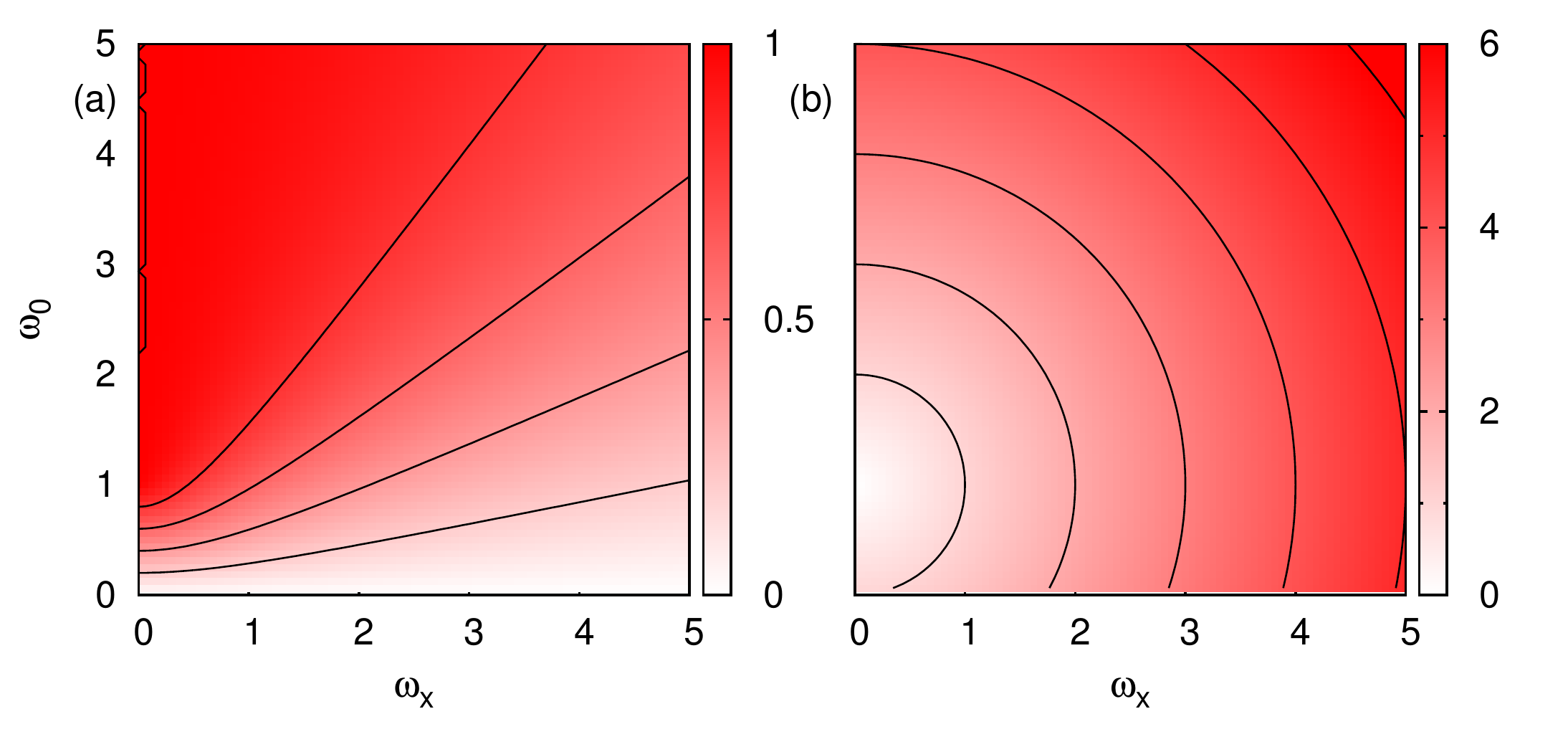}
\caption{a)1LL energy based on Eq.(\ref{omega1}). b). energy difference between Eq.(\ref{omega2}) and Eq.(\ref{omega1}). One should note that while $V_{\vec q}^{\left(10\right)}$ always corresponds to the effective interaction in the 1LL, $V_{\vec q}^{\left(01\right)}$ can correspond to very high LLs.}
\label{LLenergy}
\end{figure}

The effective interaction for the first excited LL is given by $V_{\vec q}^{\left(10\right)}$, with the energy of $\omega_1$ as given in Eq.(\ref{omega1}). One should note that with a non-zero parallel magnetic field $\omega_1<1$, so the excitation energy from LLL to 1LL is always smaller than the cyclotron energy $\omega_z$; it approaches $1$ with vanishing $\omega_x$ and $\omega_0>1$. For $\omega_0<1$, this corresponds to thick sample (relative to the strength of the perpendicular magnetic field), and $\omega_1\rightarrow\omega_0$ as $\omega_x\rightarrow 0$. The contour plot of the energy of 1LL is shown in Fig. \ref{LLenergy}(a).

\begin{figure}[htb]
\includegraphics[width=\linewidth]{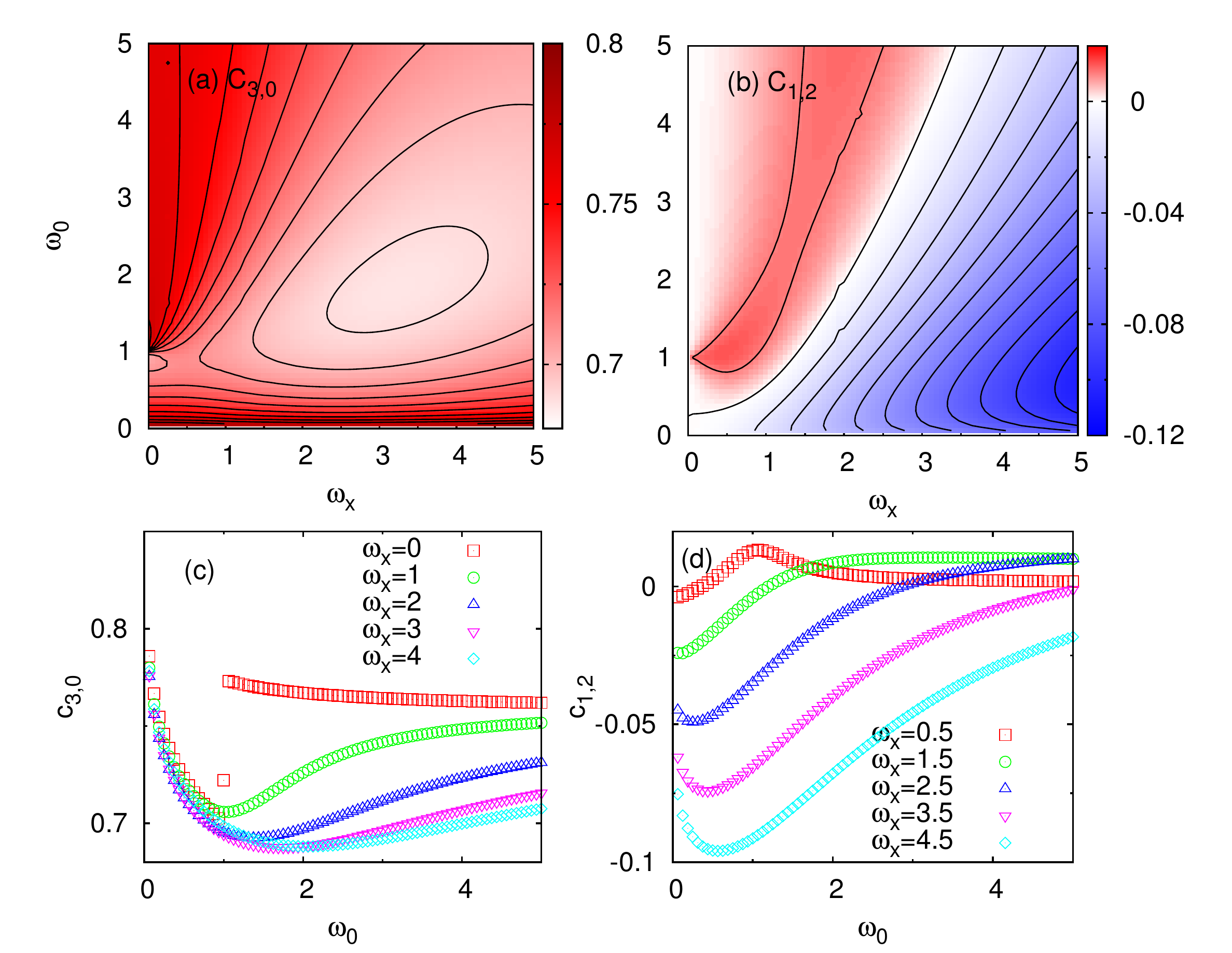}
\caption{a)The contour plot of $c^\eta_{3,0}$; b) The contour plot of $c^{\eta+}_{1,2}$; c) The dependence of $c^\eta_{3,0}$ on $m$ at different sample thickness and parallel magnetic field; d) The dependence of $c^{\eta+}_{1,2}$ on $m$ at different sample thickness and parallel magnetic field;}
\label{f4}
\end{figure}

The decomposition of $V_{\vec q}^{\left(10\right)}$ into generalized PPs has more interesting dependence on the sample thickness and the parallel magnetic field, as one can see from Fig. \ref{f4} for the expansion of the PPs in the lab frame. To understand this, it is useful to first look at the case where $\omega_x=0$. The 1LL is qualitatively different for $\omega_0<1$ and $\omega_0>1$, since for $\omega_0<1$ it is essentially a subband excitation with the form factor qualitatively different from the case when $\omega_0>1$. This explains the sharp change of $c_{m0}$ at $\omega_0=1$ in Fig. \ref{f4}(c).   

For $\omega_x>0$, the parallel magnetic field mixes the dynamics between the x-y plane and that along the z-axis, and the transition at $\omega_0=1$ becomes smooth, as one can see from Fig. \ref{f4}(c),(d). Increasing $\omega_x$ generally enhances the anisotropic PP components, but reduces the isotropic PP components. While for $\omega_x=0$, increasing the sample thickness enhances $V^\eta_1$ and thus makes the Laughlin state at $\nu=1/3$ more stable in the 1LL, that is not always the case when $\omega_x>0$. This can be clearly seen in the minima of plots in Fig. \ref{f4}(c), due to the mixing of the LLL characteristics in higher LLs.  
\begin{figure}[htb]
\includegraphics[width=\linewidth]{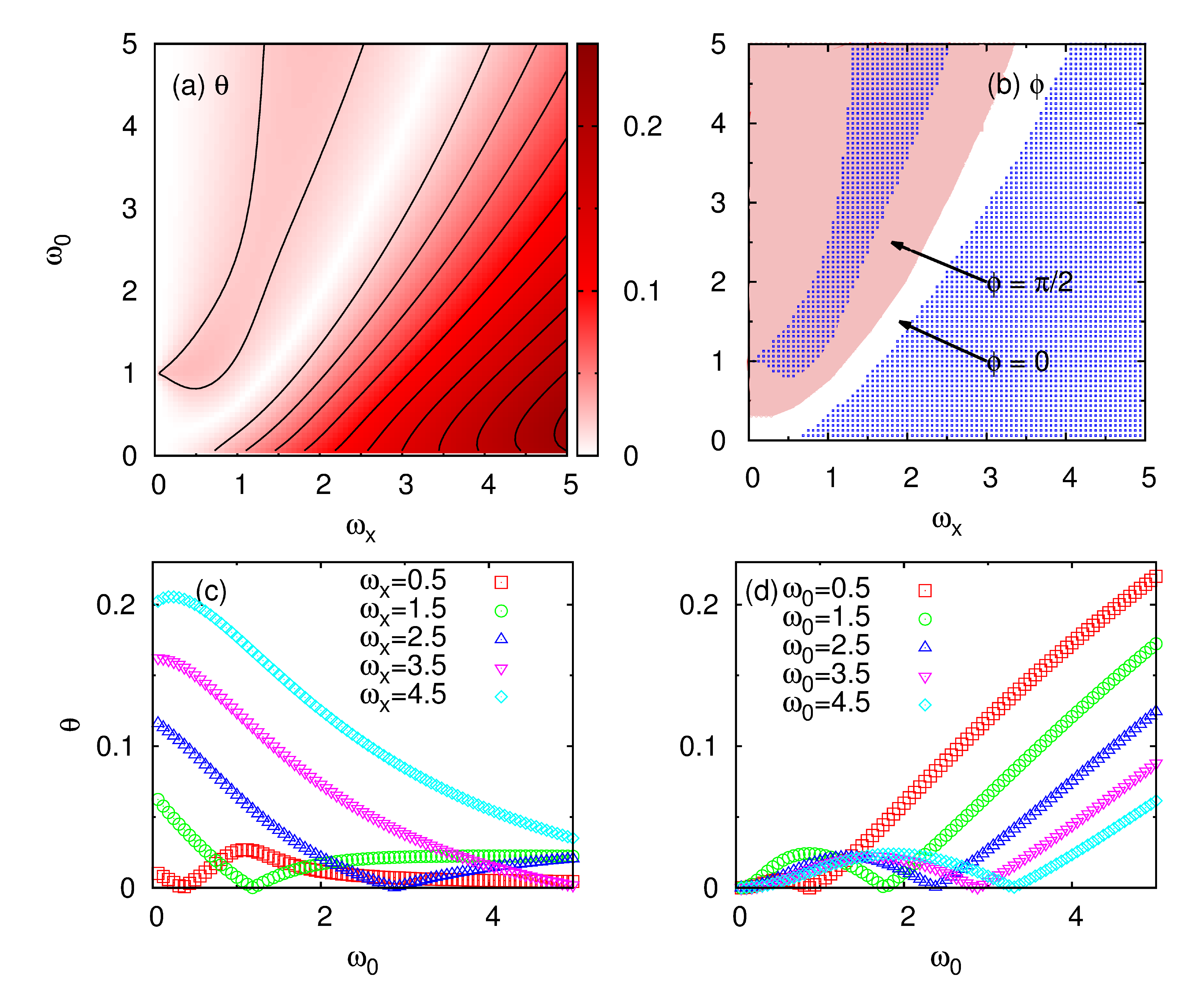}
\caption{a). The contour plot of $\theta$ of the metric that maximises $V_1$ in 1LL; b). The contour plot of $\phi$, the angle of the orientation of the metric that maximises $V_1$ in 1LL. Due to inversion symmetry $\phi$ only takes the value of $0$ and $\pi/2$, which shifts abruptly from the white region to the red region. The blue-shaded region is where $\theta>1.04$.}
\label{1LLmetric}
\end{figure}   

It is also interesting to see that $c^{\eta +}_{12}$ changes sign as the thickness increases in Fig. \ref{f4}(d), indicating a change of orientation of the effective interaction anisotropy. To observe that from a different perspective, we again look at the metric $g$ of the pseudopotential expansion that maximizes the coefficient of $V^g_1$, as plotted in Fig. \ref{1LLmetric}. Note that apart from the rich dependence of $\theta$ (which parameterizes the stretching of the metric) on $\omega_0$ and $\omega_x$, there is also a non-trivial dependence of $\phi$, which parameterizes the orientation of the stretched metric, on $\omega_x$ and $\omega_0$. One should note that due to the inversion symmetry $V^{\left(mn\right)}_{q_x,q_y}=V^{\left(mn\right)}_{q_x,-q_y}$, the metric $g$ at which $c^{g+}_{12}$ vanishes will have either $\phi=0$ or $\pi/2$. There is thus a sharp transition of the angle of the intrinsic metric of the Laughlin state (if the system is gapped) at the boundary demarcating the red and white regions in Fig. \ref{1LLmetric} (b), when the angle $\phi$ goes from $\pi/2$ to $0$.

Another feature of the metric $g$ that maximizes the coefficient of $V_1^g$ is that $\theta$ becomes vanishingly small not only at small $\omega_x$, but also close to the transition boundary of $\phi$. In this region $\phi$ is not very well-defined. On the other hand, in regions where $\theta$ is significant (the blue-shaded region in Fig. \ref{1LLmetric}(b), as long as the spectrum is gapped, the change of anisotropy orientation of the generalized Laughlin state can be observed experimentally with measurements related to ground state structure factor and the neutral excitation gap\cite{qiu11,zlatko,yang1}.
\begin{figure}[htb]
\includegraphics[width=\linewidth]{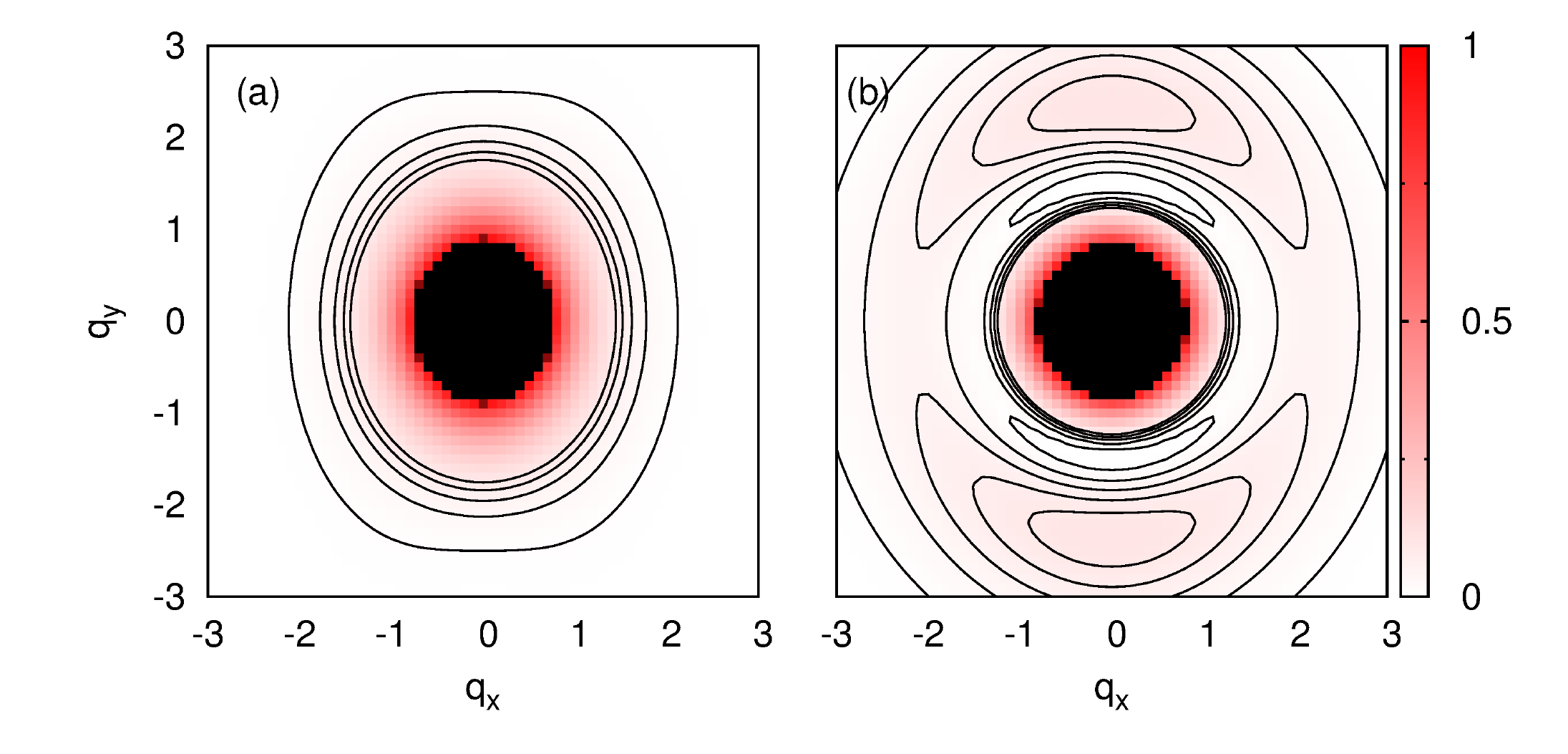}
\caption{a)$\omega_0=0.5,\omega_x=2$, the contour is squeezed along the x-axis everywhere; b) $\omega_0=4,\omega_x=2$, the contour is squeezed along the x-axis for large $|q|$, but squeezed along y-axis for small $|q|$.}
\label{contour_1LL}
\end{figure} 

The origin of such change in anisotropy orientation as one tunes $\omega_0$ (even with fixed $\omega_x$) can be illustrated by the contour plot of the effective two-body interactions (see Fig. \ref{contour_1LL}). While for large $|q|$, the anisotropy orientation of the contour lines are the same given that the direction of the parallel magnetic field is always along the x-axis, for short range interaction at small $|q|$, the orientation depends on the details of the effective interaction. Since for Laughlin state at $\nu=1/3$, only short range interaction is important, and the maximization of the coefficient of $V^g_1$ is equivalent to the vanishing of $c_{1,2}^{g\pm}$, the change of anisotropy orientation can be observed as long as the gap for this Laughlin state is robust. One can also conjecture that for other topological states where longer range interaction (involving $c^{g\pm}_{mn}$ with larger $m$) are important, the change of orientation will no longer be present for the incompressible states.

\section{Applications and Discussions}\label{tlimit}

In this section, we discuss some of the more specific applications of the detailed analysis in the previous sections, from both the theoretical and experimental point of view. The theoretical discussions concerning the computation of the emergent metric of the FQH states in the thermodynamic limit are not constrained to the anisotropic 2DEG system with in-plane magnetic field, and because of that we illustrate the main idea with a simpler and commonly studied model with anisotropic band mass tensor. The second part of the section focuses on some of the past experiments on tilted magnetic field measurements of various compressible and incompressible states, in which certain interesting experimental observations can be explained by the analysis in this paper.

\subsection{Computation of Emergent Metric in Thermodynamic Limit}

Since Haldane pointed out the geometric degree of freedom in fractional quantum Hall fluids\cite{qiu11,haldane3,haldane4}, it has been the common practice to numerically extract the emergent metric of the topological ground state from finite size systems, via exact diagonalization followed by wavefunction overlap, or with ground state energy minimization variationally\cite{yang1,zlatko,regnault}. The variation of the metric in the expansion of the generalized PPs, on the other hand, allows us to compute this metric analytically. As shown in the previous section, one can maximize the coefficient of $V_1$, and the metric at which $V_1$ is maximized also minimizes the perturbation to the generalized Laughlin state thanks to Eq.(\ref{identity}). This analytically obtained metric is the emergent metric of the fractional quantum Hall fluid in the \emph{thermodynamic limit}, free of the constraints of the finite size and boundary effects of numerical analysis. 

In general, the analytic results show that the emergent metric is less anisotropic than the numerical calculation of the largest system size attainable would suggest. This is obvious if we compare the results of Fig.\ref{LLLtheta}, Fig. \ref{1LLmetric} and those in Ref. \onlinecite{zlatko}. To illustrate this in more details, we look at a simple but commonly used effective two-body interaction as follows:
\begin{eqnarray}
&&V_{\vec q}=\frac{1}{|q|_c}\left(F_N\left(|q|_m\right)\right)^2\label{simplev}\\
&&|q|_c=\left(q_x^2+q_y^2\right)^{\frac{1}{2}},|q|_m=\left(\alpha_m^2q_x^2+q_y^2/\alpha_m^2\right)^{\frac{1}{2}}
\end{eqnarray}

\begin{figure}[htb]
\includegraphics[width=\linewidth]{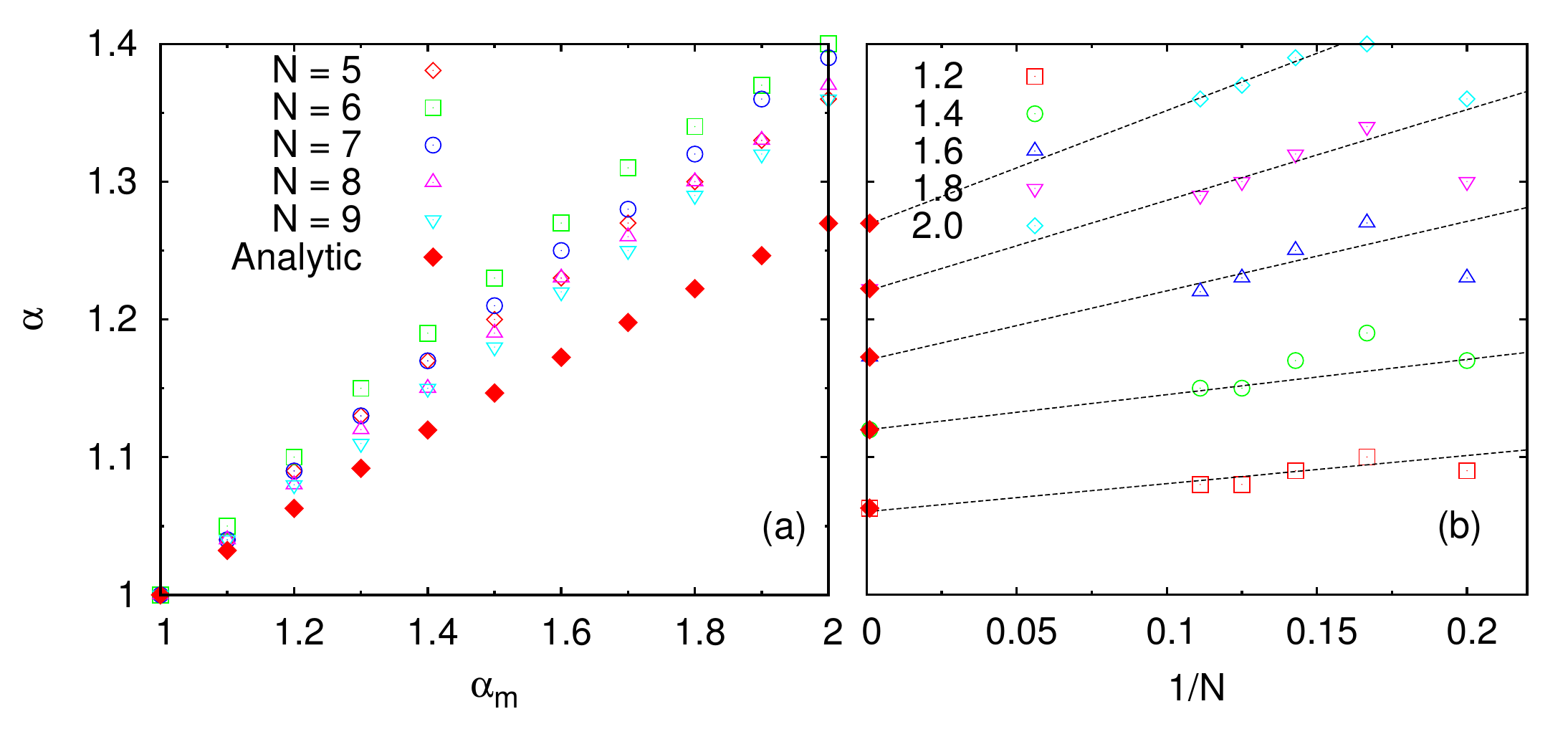}
\caption{The intrinsic metric $\alpha$ as a function of $\alpha_m$. a) Comparing of the numerical results for different systems on torus geometry with square lattice, and the analytic results in the thermodynamic limit; b). Finite size scaling and the analytic results, shown by solid dots at the y-axis. The straight lines serve as guide of eyes for the extrapolation of the numerical results to the infinite system size.}
\label{numericanalytic}
\end{figure}

This is an effective interaction with isotropic Coulomb interaction and a form factor from an anisotropic effective mass tensor given by the metric $g_m=\left(\alpha_m,0,0,1/\alpha_m\right)$. Such quantum Hall systems can be realised in samples with anisotropic band structure, stretched graphene and black phosphorous. In the LLL, the ground state at filling factor $\nu=1/3$ is the Laughlin state with an anisotropic emergent metric, which was studied with exact diagonalization and wavefunction overlap\cite{yang1}. Here we do a finite size scaling of the numerical calculation and compare it with the analytically obtained metric by expanding Eq.(\ref{simplev}) in the basis of the generalized PPs with the metric that maximizes $V_1$. The result is shown in Fig. \ref{numericanalytic}.

The comparison between the finite size numerical calculation of the intrinsic emergent metric of the Laughlin state, and the analytic computation from the generalized PPs, shows that in general, the former over-estimates the anisotropy of the ground state Laughlin states, as long as the Hamiltonian is gapped. This is especially true when the microscopic anisotropy of the effective interaction is small, which is the case both for the exact expression in Eq.(\ref{effective2d}) or the simple model in Eq.(\ref{simplev}), even when the in-plane magnetic field or the anisotropy in the band mass tensor are strong. Finite size scaling partially compensates for such over-estimation, though it is less definitive and computationally much more intensive. We conjecture this is a general feature for many experimental systems where the source of anisotropy is mainly from the single particle dynamics.

\subsection{Tilted Magnetic Field Experiments}

\begin{figure}[htb]
\includegraphics[width=\linewidth]{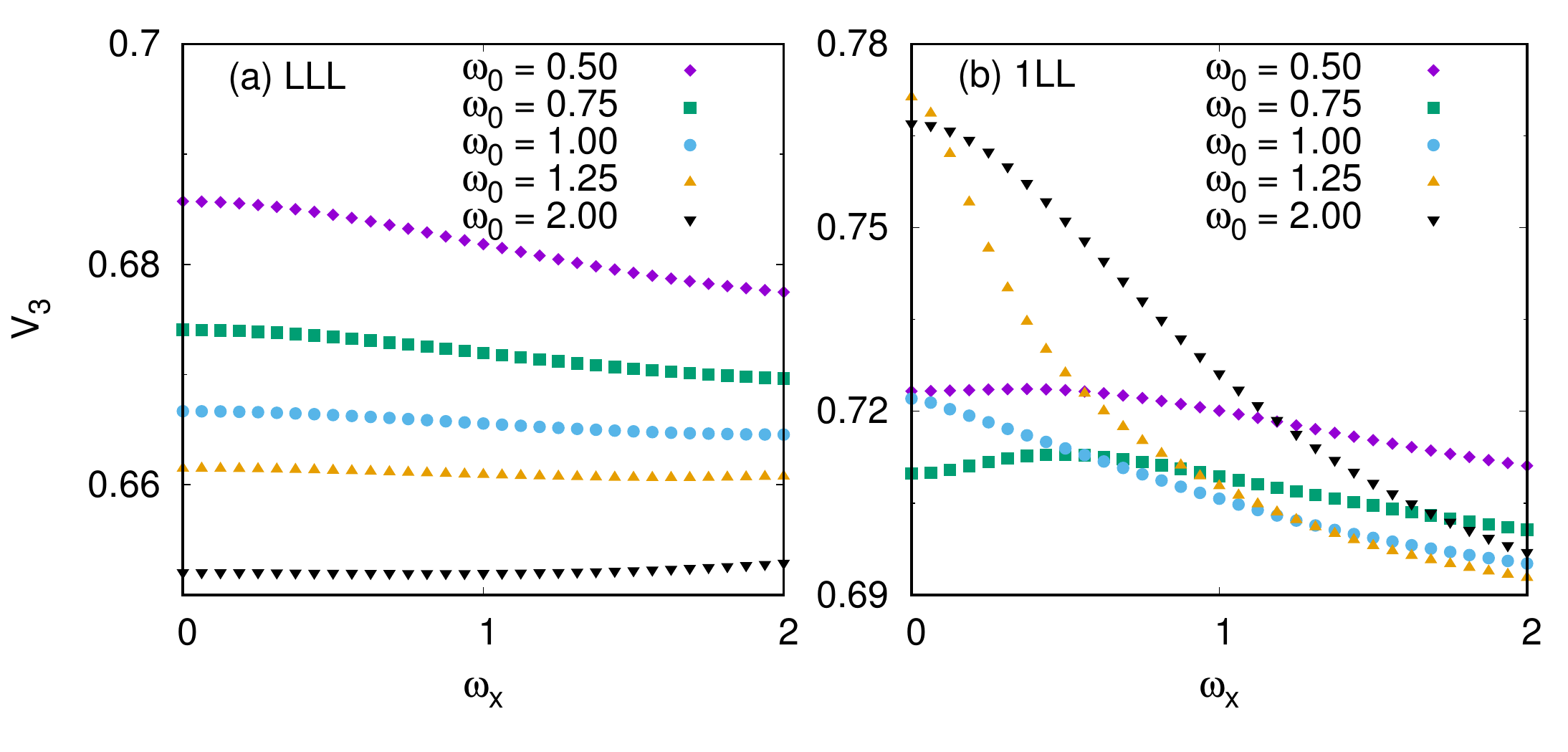}
\caption{The coefficient of $V_3$ when that of $V_1$ is normalized to $1$, at the metric that maximizes $V_1$, as a function of in-plane magnetic field at different sample thickness for a). LLL, b). 1LL.}
\label{LLLvs1LL}
\end{figure}

A number of experiments on finite thickness 2DEG systems in GaAs/AlGaAs quantum well (QW) with tilted magnetic field have been reported on higher LL ~\cite{lilly99, du99}, and those in LLL are rarely studied. The effect of anisotropy introduced by the in-plane magnetic field has been quite consistent in the LLL, especially with the Laughlin state at $\nu = 1/3$. Based on extensive numerical studies\cite{yang1,zlatko}, the incompressibility at $\nu=1/3$ is not easily destroyed by the in plane magnetic field, though the stability (as measured by the energy gap) generally decreases when the system becomes more anisotropic. These results agree with our analysis in Sec.~\ref{LLL_description}, where the coefficients of $V_3$ as well as other isotropic PPs increase with $\omega_x$ as long as the QW (or heterojunction) width is not too thick (also see Fig.(\ref{LLLvs1LL})), in addition to the introduction of anisotropic PPs. This is true for PPs with both the lab metric and the metric that maximizes $V_1$, and for the stability of the gapped Laughlin state, the latter is the more relevant measure.

In contrast, rich physical behaviors are observed in 1LL and higher LLs; results from different experiments and different samples are comparatively much less consistent. For the incompressible states, the main questions center around how the stability of the Laughlin states (at $\nu=7/3,8/3$) in 1LL, and the Moore-Read states (at $\nu=5/2,7/2$) is affected by increasing the tilted $B$-field. Compressible states such as the bubble and stripe phases are also common and of great interest in higher LLs (e.g. $\nu=9/2, 11/2$), especially for the direction of the measured anisotropy, and the transitions from incompressible phases to the compressible phases. A detailed review of these experimental results in the framework of the theoretical tools developed in this paper is work in progress and will be presented elsewhere. Here we give brief discussions and explanations for some of the most prominent experimental observations in this field.

The pseudopotential coefficients of the effective interaction in the 1LL depend very differently on the in-plane magnetic field, compared with those in the LLL (see Sec.~\ref{1LL_description} and Fig.(\ref{LLLvs1LL})). In particular for the Laughlin state, if we expand the effective interaction in the metric that maximizes the coefficient of $V_1$, the coefficients of other isotropic PPs decrease significantly with increasing $\omega_x$. Since the coefficients of anisotropic PPs are much smaller than those of the isotropic PPs, their effects on the stability of the states are small. Decreasing components of $V_3, V_5$ indicates that the stability of the Laughlin states at $\nu=7/3,8/3$ tends to be enhanced with anisotropy introduced by the in-plane $B$-field, at least for small $\omega_x$ or large $\omega_0$. The increase of the activation gaps at $\nu = 7/3, 8/3$ are indeed observed in several cases\cite{dean08,chi12, chithesis}. In a low density electron system ($n \sim 1.0 \times 10^{11}$ cm$^{-2}$) in a 50 nm wide QW, $\nu = 7/3$ (at $B_{\perp} = 1.3 $ T) states are strengthened by in-plane $B$-field ~\cite{chi12}.
These are consistent with our analysis just from the projected effective interaction itself. One should note that at the same filling factor, low electron density implies smaller perpendicular $B$-field. Since in our calculations $\omega_0$ is normalized by $\omega_z$, such experimental systems correspond to relatively large $\omega_0$, and in those cases the coefficient of $V_3$ (as well as for other isotropic pseudopotentials) decreases significantly with increasing $\omega_x$, enhancing the stability of the incompressible state. This is in contrast to some results from the earlier report ~\cite{peterson08} where finite size numerical computations were performed, and the in-plane $B$-field is only implicitly accounted for with the reduction of the effective thickness.
On the other hand, for $\omega_x=0$ our results agree with most conclusions from the same work\cite{peterson08}.

In a high density electron system ($n \sim 6 \times 10^{11}$ cm$^{-2}$) in a narrow (20 nm) QW, the gaps of 5/2 and 7/3 states exhibit the similar behaviors: $\nu = 7/3$ (at $B_{\perp}=11$ T) decrease gradually with tilted $B$-fields (or total $B$-field) ~\cite{chi10}. This corresponds to the cases where $\omega_0$ is relatively small, and much slower decrease of the coefficient of $V_3$ (even with initial increase at small $\omega_x$) can be seen in Fig.(\ref{LLLvs1LL}b). In another experiment on low density ($n \sim 1.5 \times 10^{11}$ cm$^{-2}$) in a 40 nm wide QW, the activation gap for the Laughlin states in 1LL is observed to decrease more dramatically under stronger in-plane $B$-fields, and it is suggested that the QW width play an important role ~\cite{dean08}. Comparing to other experimental systems ~\cite{chi12}, the electron density is higher by $50\%$ while the sample thickness is smaller by $20\%$, so by a rough estimate the effective $\omega_0$ here\cite{dean08} is still smaller.
While the experimental reports ~\cite{chi10} follow the traces of low value $\omega_{0}$, and the results for the low density electron system ~\cite{chi12} belongs to the case of large $\omega_{0}$ (Fig.(\ref{LLLvs1LL}b)), detailed analysis on the actual profile of the quantum well confinement and the competition between decreasing isotropic PPs versus increasing anisotropic PPs is needed for bouderline cases\cite{dean08}.

More accurate theoretical predictions thus require the incorporation of various experimental details, including the confining potentials and the implicit tuning of the carrier densities. While in this paper the harmonic confinement well is used for its analytic tractability, realistic confinement can lead to different effective width of the quantum well. At fixed filling factors, different perpendicular magnetic fields also correspond to different carrier densities, which in turn also affect the effective thickness of the sample. For higher LLs, the mixing of the cyclotron levels and the subbands also can lead to strong LL mixing, which may need be computed perturbatively with the appropriate intrinsic metric even for the Laughlin states.

Many experiments also focus on the more interesting Moore-Read (MR) states in the 1LL. In general if the system is incompressible, the stability of the MR states decreases with stronger in-plane magnetic field ~\cite{dean08,chi10}.
Exceptions are also observed where the stability of the MR state strengthens with the in-plane field ~\cite{chi12, chithesis}, and based on the experimental conditions, the enhancement of the activation gap could be mostly due to a more spin-polarized ground state. While the analysis in this paper focuses more in the Laughlin states, a microscopic understanding of the stability of the spin-polarized MR states is also possible in principle. One should note that the MR states are much more complicated, because with only two-body interactions it is not entirely clear which combination of the pseudopotentials gives the most stable state (unlike the Laughlin state, which is the exact ground state of $V_1$). The incorporation of the three-body interactions from LL mixing for the anisotropic systems are thus a necessity in predicting the optimal experimental conditions and optimal engineering of the interaction. The intrinsic geometric degrees of freedom for the MR model Hamiltonian are also much more non-trivial as compared to its two-body counterpart, because of the possibility of a four-dimensional Bogoliubov transform in the Hamiltonian ~\cite{yang_future}.

For the compressible half-filling states in 2DES, especially for those at $\nu=9/2, 11/2,...$ at higher LL, experimental results mainly focus on the longitudinal resistances of different crystalline orientations, namely $R_{xx}$ and $R_{yy}$ ~\cite{lilly99, du99, lilly99_2, chithesis}. Such anisotropic phenomena can be understood as the stripe phase: under a perpendicular magnetic field, the stripes are along the [110] direction; however, there are exceptions in experiments ~\cite{zhu02}.
Therefore, the native symmetry breaking from the details of the conductance or valence band can be affected by the carrier density (or $B$-field). On the other hand, in-plane magnetic field provides an external symmetry breaking, which competes with native symmetry breaking, such as those in the reports ~\cite{lilly99_2, pan99, chi10}.
Possibly, the native symmetry breaking comes from the intrinsic band mass anisotropy, which is independent to the tilting of the magnetic field. Only when the band mass anisotropy is aligned with the anisotropy of the in-plane magnetic field, the anisotropy measurement of the longitudinal resistances can be consistent and easily predicted.
When they are not aligned, the competition between the native symmetry breaking and $B_{\parallel}$-induced symmetry breaking can be complicated. Such complications can also be qualitatively different in different LLs ~\cite{shi16}, when the form factors have different contributions from the subband wavefunctions and cyclotron wavefunctions. 

For these cases, as the experimental measurements are based on the lab metric, we can compute the anisotropy of the effective interactions in the language of PPs with the lab metric unambiguously for the compressible states, incorporating all microscopic details of the experimental systems. A detailed study of the relevant experiments will also be presented elsewhere.

\section{Conclusions}\label{summary}

In this paper, we model the quantum Hall sample with a finite thickness with a harmonic well confining potential in the direction perpendicular to the Hall surface, and analytically calculated the effective two-body interaction between electrons within a single LL, when both perpendicular and in-plane magnetic fields are present. The analytical results and small parameter expansion show that using a band mass anisotropy to model the in-plane magnetic field is generally a crude approximation, especially in 1LL or higher LLs. The characterization of such interaction in the language of generalized PPs show that the anisotropy induced by tilting the magnetic field is generally quite small both within LLL and 1LL. On the other hand, both the sample thickness and the strength of the in-plane magnetic field can be used experimentally to effectively tune the pseudopotentials, for the stabilization of incompressible topological states and for the transition between different phases of the quantum Hall fluids. 

For anisotropic quantum Hall systems, one should note there is no preferred metric microscopically, but when the system has a gap in the thermodynamic limit, there is a well-defined emergent metric one can compute by varying the metric in the generalized PP expansion. For anisotropic Laughlin states at $\nu=1/3$ realized in such quantum Hall systems, we compute analytically its intrinsic emergent metric in the thermodynamic limit, which previously can only be estimated with numerical computations of finite size systems. Our results reveal an interesting reorientation of the anisotropy direction in the 1LL, at certain thickness range when we increase the strength of the in-plane magnetic field but not its direction.  This emergent metric is the preferred metric for the experimentally measurable ground state properties (e.g. ground state structure factors). For perturbative calculations and estimating the robustness and energy gap of the incompressible state, one should look at the expansion of the realistic Hamiltonian in the PP basis in this emergent metric. Such expansion is useful for guiding the fabrication of samples with specific parameters, especially when exact diagonalization of the complicated effective two-body interaction becomes inconclusive due to finite size effect.

We have also discussed various experiments of the FQH systems with in-plane magnetic field. For theoretical understandings of the complicated effective interactions with different experimental parameters, it is important to expand the effective interaction in the generalized PP basis with appropriate metric. This allows us to explain the contrasting behaviours of the Laughlin state stability in different LLs, when the strength of the in-plane magnetic field is tuned. One should note that our analytical calculation in this paper assumes harmonic confining potential along the vertical direction, with a single length scale characterizing the width of the well. In reality, the actual confining potential is rather complicated and dependent on the electron density; thus the subband energies are no longer evenly spaced and the effective potentials are not calculable analytically. Nevertheless, we expect many qualitative behaviours can be captured by the model we used in this paper, where the realistic quantum well can be characterized by an effective width\cite{zlatko}. Moreover, we can in principle take into account of all experimental details (including the realistic quantum well profile and the lattice based effective mass tensor) for the numerical computation of the effective interactions in different LLs. The formalism of the generalized PPs can be applied generically, to accurately predict many interesting experimental features at different filling factors in higher LLs.

In realistic experimental systems, it is quite often that the in-plane magnetic field is not the only source of anisotropy. For example, in systems with band mass anisotropy, the anisotropic effective mass tensor is an additional source of anisotropy~\cite{yang1}. From the numerical perspective, we can easily model this on the torus with non-trivial aspect ratio between two principle periodic directions, and analyse the effective two-body interactions with generalised PPs in the usual way. Many interesting phase transitions in the 1LL can be related to the interplay and competition between these two different sources of anisotropy. It is also of particular interest to understand how the thin torus limit~\cite{Tao, Rezayi94, Bergholtz} can be used to characterize the anisotropic topological states. For the isotropic cases, the ground states of the thin torus limit leads to the ``root configurations" of the model wavefunctions in the isotropic cases; when rotational symmetry is broken, such connections are no longer obvious, and a systematic generalization is needed. We will study these interesting issues elsewhere.

The methodology we presented in this paper can also be generalized to more exotic fractional quantum Hall fluids, such as the Moore-Read states when we compute the LL mixing. Similar in these cases, when rotational symmetry is broken (the effective interaction is not characterized by a unique metric), the stability of the incompressible quantum Hall fluids cannot be determined by expansion of the effective interaction into the pseudopotentials with the cartesian metric (the metric of the lab frame) or any other arbitrary metric. For two-body interactions, if the expansion of the anisotropic interaction into the PPs of the cartesian metric has a very small $V_1$ component, this does not imply that the Laughlin state is not supported. If we expand the interaction with the metric that maximises the coefficient of $V_1$, and in that metric $V_1$ is the dominant component, we will still have an incompressible Laughlin state with a non-trivial metric. Conceptually the same is true for three-body interactions, though the metric degree of freedom in three-body pseudopotentials is physically much more non-trivial than the two-body cases\cite{yang_future}. Understanding such geometric aspects of effective interactions can be important in helping to realize more robust non-abelian quantum Hall fluids, and we will discuss this issue in more details elsewhere.

\begin{acknowledgements}
We thank Zlatko Papic for useful discussions. This work was supported in part by Singapore A*STAR SERC ``Complex Systems" Research Program grant 1224504056. Z.-X.H. is supported by National Natural Science Foundation of China Grants No. 11674041, 91630205, Fundamental Research Funds for the Central Universities Grant No. CQDXWL-2014-Z006 and Chongqing Research Program of Basic Research and Frontier Technology Grant No. cstc2017jcyJAX0084. C. Z. is supported by the National Science Foundation of China (Grant No.11374020, 11674006).
\end{acknowledgements}

\end{document}